\documentclass[aip,graphicx,pop,preprint]{revtex4-1}

\usepackage{graphicx}
\usepackage{amsmath}
\usepackage{subcaption}
\usepackage{xcolor}

\draft % marks overfull lines with a black rule on the right

\begin{document}

%\preprint{}

\title{Verification of the global gyrokinetic stellarator code XGC-S for linear ion temperature gradient driven modes}



\author{M.~D.~J.~Cole}
\email[]{mcole@pppl.gov}
\author{R.~Hager}
\affiliation{Princeton Plasma Physics Laboratory,\\
Princeton University, Princeton, New Jersey 08543, USA}

\author{T.~Moritaka}
\affiliation{National Institute for Fusion Science,\\
Toki 509-5292, Japan}

\author{J.~Dominski}
\affiliation{Princeton Plasma Physics Laboratory,\\
Princeton University, Princeton, New Jersey 08543, USA}

\author{R.~Kleiber}
\affiliation{Max Planck Institute for Plasma Physics,\\
D-17491 Greifswald, Germany}

\author{S.~Ku}
\affiliation{Princeton Plasma Physics Laboratory,\\
Princeton University, Princeton, New Jersey 08543, USA}

\author{S.~Lazerson}
\affiliation{Princeton Plasma Physics Laboratory,\\
Princeton University, Princeton, New Jersey 08543, USA}

\author{J.~Riemann}
\affiliation{Max Planck Institute for Plasma Physics,\\
D-17491 Greifswald, Germany}

\author{C.~S.~Chang}
\affiliation{Princeton Plasma Physics Laboratory,\\
Princeton University, Princeton, New Jersey 08543, USA}

\date{\today}

\begin{abstract}
XGC (X-point Gyrokinetic Code) is a whole-volume, total-$f$ gyrokinetic particle-in-cell code developed for modelling tokamaks.
In recent work, XGC has been extended to model more general 3D toroidal magnetic configurations, such as stellarators.
These improvements have resulted in the XGC-S version.
In this paper, XGC-S is benchmarked for linear electrostatic ion temperature gradient-driven microinstabilities, which can underlie turbulent transport in stellarators.
An initial benchmark of XGC-S in tokamak geometry shows good agreement with the XGC1, ORB5, and global GENE codes.
A benchmark between XGC-S and the EUTERPE global gyrokinetic code for stellarators has also been performed, this time in geometry of the optimised stellarator Wendelstein 7-X.
Good agreement has been found for the mode number spectrum, mode structure, and growth rate.
\end{abstract}


\maketitle %\maketitle must follow title, authors, abstract and \pacs

\section{Introduction}
\label{introduction}

Anomalous transport, believed to be principally turbulent, is known to be the main cause of heat and particle loss in high performance tokamaks.
Anomalous transport has also been observed in stellarator experiments, including Wendelstein 7-AS~\cite{stroth98}, Large Helical Device (LHD)~\cite{motojima03} and Helically Symmetric eXperiment~\cite{deng15}.
In the second operations phase of the Wendelstein 7-X (W7-X) stellarator~\cite{nuehrenberg86,wolf17} (OP 1.2), turbulence has been observed to be the dominant driver of transport~\cite{klinger19}.
Theoretical work has shown that turbulent transport in stellarators can be qualitatively different to that in tokamaks~\cite{proll12,xanthopoulos14,plunk19}.
Understanding turbulent transport in stellarators is therefore increasingly relevant.

Numerical tools are useful for investigating turbulent transport.
One of the most effective physical models for numerical modelling of plasma turbulence has been 5D gyrokinetics, which combines high physical completeness with reduced computational resource requirements compared to a 6D kinetic model.
Gyrokinetics is an approximation to the Vlasov-Poisson (or Vlasov-Maxwell) system that is valid for phenomena of low frequency relative to the ion gyrofrequency~\cite{hahm07}.
The fast gyromotion is decoupled from the slower guiding centre motion.
The 6D Vlasov equation is thus reduced to a 5D gyrokinetic equation.

In stellarator geometry, gyrokinetic simulation of ion-scale microinstabilities has been carried out with local models~\cite{proll13}, flux-surface global models~\cite{xanthopoulos14,xanthopoulos16}, and models fully global to the last closed flux surface~\cite{helander15,riemann16}.
In the former two categories, nonlinear simulations have been performed to calculate saturated turbulent states and fluxes.
Microinstability calculations global to the last closed flux surface have only been demonstrated with the EUTERPE code, linearly with a delta-$f$ model~\cite{kornilov04,helander15,riemann16}.
Neoclassical transport has been modelled with a global full-$f$ gyrokinetic model to the last closed flux surface with the GT5D code~\cite{matsuoka18}.
The GTC~\cite{spong17} and MEGA~\cite{todo17} codes have also been used to perform global simulations to the last closed flux surface in LHD geometry.
In these codes, an energetic particle species has been treated with a gyrokinetic or drift kinetic model, interacting with a background plasma red treated as an MHD fluid in MEGA and with a gyrokinetic ion/fluid electron hybrid model in GTC.

The gyrokinetic total-$f$ code XGC~\cite{ku09,ku15,ku18} is global to the first wall, and has so far been applied to tokamaks including the pedestal and scape-off layer in the simulation volume~\cite{churchill17, chang17}.
Recently, XGC has been generalized to 3-dimensional geometry for application to stellarator plasmas~\cite{moritaka_submitted,cole19}.
In the present paper, XGC-S is verified for linear Ion Temperature Gradient-driven (ITG) microinstabilities with a delta-$f$ model global to the last closed flux surface in axisymmetric and stellarator geometries.
In Section~\ref{model_verification}, the model is described and simulation results are presented for global ITG calculations with a standard circular tokamak case.
This case has been used for benchmarking between the tokamak XGC1 code and other gyrokinetic codes.
XGC-S correctly reproduces results of this benchmark.
In Section~\ref{w7x_results}, the analysis is extended to linear global ITG instabilities in the Wendelstein 7-X (W7-X) stellarator~\cite{nuehrenberg86,wolf17}, in comparison with the EUTERPE code~\cite{helander15, riemann16}.
We conclude in Section~\ref{conclusion}.

\section{Model verification}
\label{model_verification}

\subsection{Model description}

XGC is a whole volume global total-$f$~\cite{ku15,ku18} gyrokinetic Particle-in-Cell (PIC) code.
Its nonlinear tokamak version for neoclassical and turbulent physics is XGC1.
It discretises the distribution functions of plasma species by numerical marker particles advanced in cylindrical ($R$, $\varphi$, $Z$) coordinates according to the gyrocenter equations of motion.
Charge and current densities are accumulated on nodes of unstructured meshes.
Potentials are calculated on mesh nodes, and fields are then calculated by taking spatial derivatives of the potentials.
These are then interpolated for the marker particle equations of motion.
A fully nonlinear Fokker-Planck Landau operator~\cite{hager16} is used to model Coulomb collisions and a Monte-Carlo method is used for neutral recycling and transport with atomic interactions~\cite{stotler17}.

XGC-S does not yet include all of these features.
In this paper, the conventional collisionless delta-$f$ method is used in electrostatic mode, with a single thermal ion species and the adiabatic approxmation for electrons.
In the delta-$f$ formulation, $f=f_{0}+\delta f$, where $f$ is the total distribution function, $f_{0}$ is a prescribed background distribution function, and $\delta f$ is a small perturbation to this background represented by the markers.
In this paper the background distribution function is always Maxwellian.

Neglecting collisions and any background electric field, $\phi_0$, the perturbed distribution function evolution equation~\cite{hahm07} solved by XGC-S is
\begin{equation}
\frac{\partial~\delta f}{\partial t} + \dot{\vec{X}} \cdot \frac{\partial~\delta f}{\partial \vec{X}} + \dot{v}_{\|} \frac{\partial~\delta f}{\partial v_{\|}}
 = -\dot{\vec{X}}_1 \cdot \frac{\partial f_0}{\partial \vec{X}} - \dot{v}_{\|1} \frac{\partial f_0}{\partial v_{\|}}
\end{equation}
where $\vec{X}$ is the gyrocentre position,
\begin{align}
\dot{\vec{X}} = \frac{1}{G}\left[v_{\|}\vec{b}+\frac{mv_{\|}^{2}}{qB}\nabla\times\vec{b} + \frac{1}{qB^2}\vec{B}\times\left(\mu\nabla B +q\nabla\langle\phi\rangle\right)\right],
\end{align}
$v_{\|}$ is the gyrocentre velocity,
\begin{equation}
\dot{v}_{\|} = - \frac{1}{mG}\left(\vec{b}+\frac{mv_{\|}}{qB}\nabla\times \vec{b}\right) \cdot \left( \mu \nabla B + q \nabla \langle\phi\rangle \right)
\end{equation}
and
\begin{equation}
G = 1 + \frac{mv_{\|}}{qB}\vec{b}\cdot\left(\nabla\times \vec{b}\right),
\end{equation}
with $B$ being the local magnitude of the magnetic field, $q$ the ion charge, and $\mu$ the magnetic moment.
$\langle\rangle$ represents the gyroaverage operation.
The corresponding $\dot{\vec{X}}_0$ and $\dot{v}_{\|0}$ are as above but excluding terms containing $\phi$, while the corresponding $\dot{\vec{X}}_1$ and $\dot{v}_{\|1}$ contain only these terms.

The EUTERPE simulations with which XGC-S is compared in this paper have been run excluding nonlinear terms from the equations of motion. 
That is, $\dot{\vec{X}}_0$ and $\dot{v}_{\|0}$, without electrostatic perturbations, are used in place of $\dot{\vec{X}}$ and $\dot{v}_{\|}$ when acting on the derivatives of the perturbed distribution function.
Since the initial field perturbations used with XGC are small, and the simulations are run for linear physics while the perturbations are small, the two models are very closely approximately equivalent in the simulations performed for this paper.

To solve these equations, the background magnetic field, $\vec{B}$, must be known at all positions.
The components of the background magnetic field, and the flux label $s=\Psi/\Psi_{LCFS}$ where $\Psi$ is the poloidal flux, are obtained from the VMEC code.
VMEC is an ideal MHD equilibrium code which assumes nested flux surfaces~\cite{hirshman83}.
The VMEC output is mapped to cylindrical coordinates for use with XGC.
To simulate linear ITG modes, the perturbed electrostatic potential must be calculated.
The gyrokinetic Poisson equation in the Pad\'e approximation is 
\begin{equation}
-\nabla_{\perp}\cdot\frac{\rho_i^2}{\lambda_{Di}^2}\nabla_{\perp}\phi=e\left(1-\nabla_{\perp}\cdot\rho^2_i\nabla_{\perp}\right)\left(\langle\tilde{n}_i\rangle-\tilde{n}_e\right),
\end{equation}
where $\rho_i$ is the ion gyroradius, $\lambda_{Di}$ is the ion Debye length, and $\tilde{n}_s$ are the perturbed species densities.
In this paper, the perturbed electron density $\tilde{n}_e$ is obtained by use of the adiabatic electron approximation.
The Pad\'e approximation is used in this section, while the short wavelength term $\nabla_{\perp}\cdot\rho^2_i\nabla_{\perp}$ is set to zero (long wavelength approximation) is used for Wendelstein 7-X simulations in the following section to permit comparison with published results of the EUTERPE code, obtained with the long wavelength approximation.

\begin{figure}
\includegraphics[origin=c,scale=0.2]{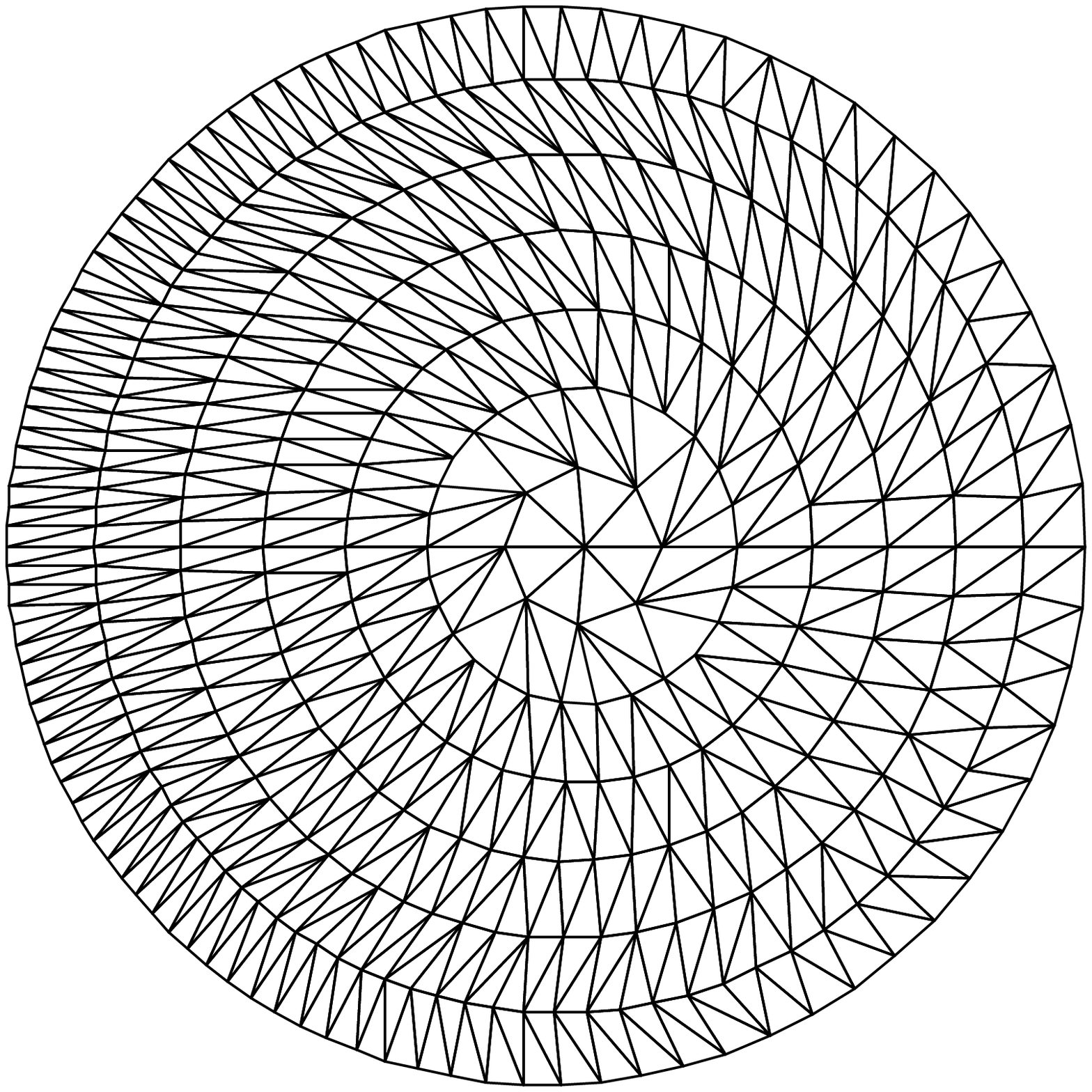}
\includegraphics[origin=c,scale=0.2]{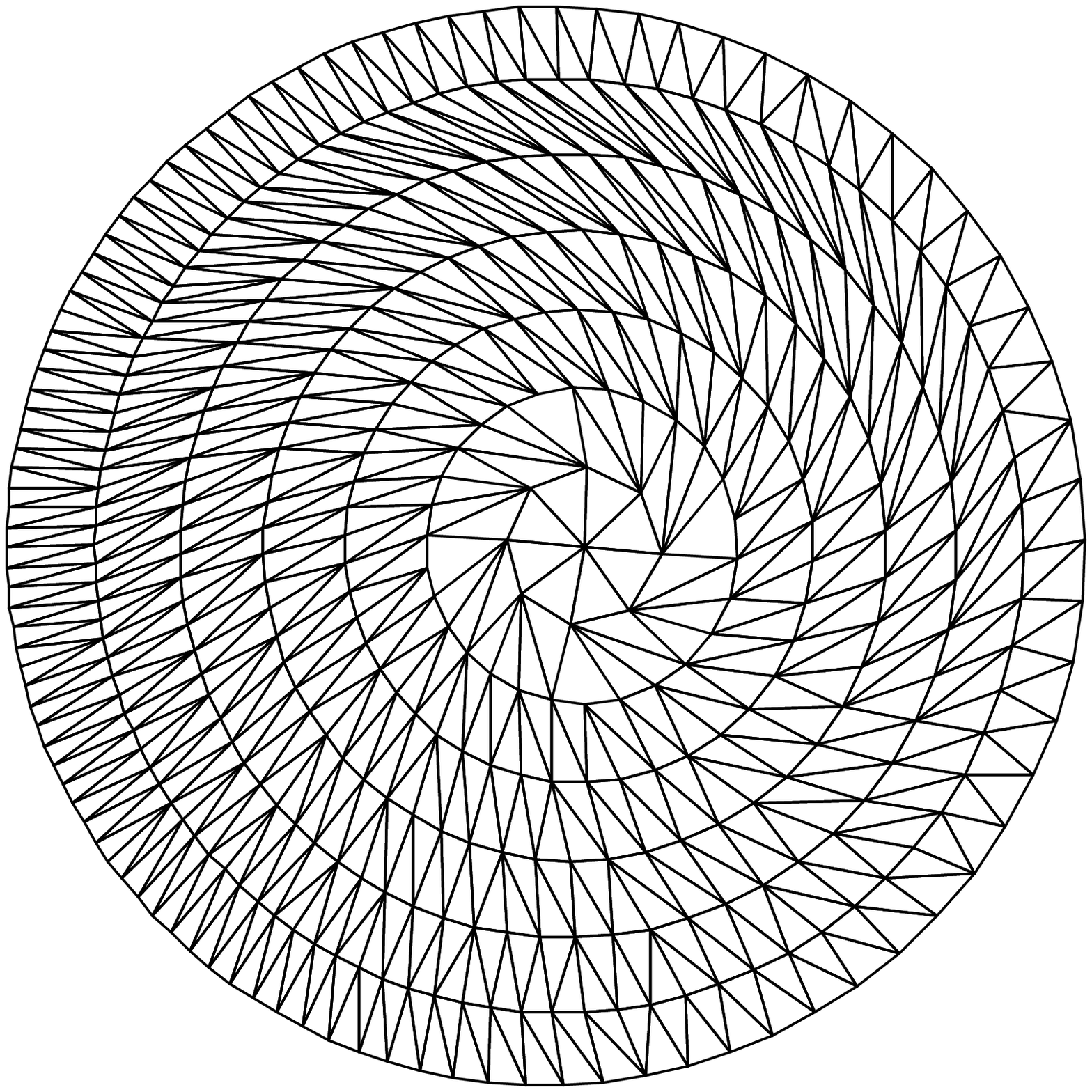}
\includegraphics[origin=c,scale=0.2]{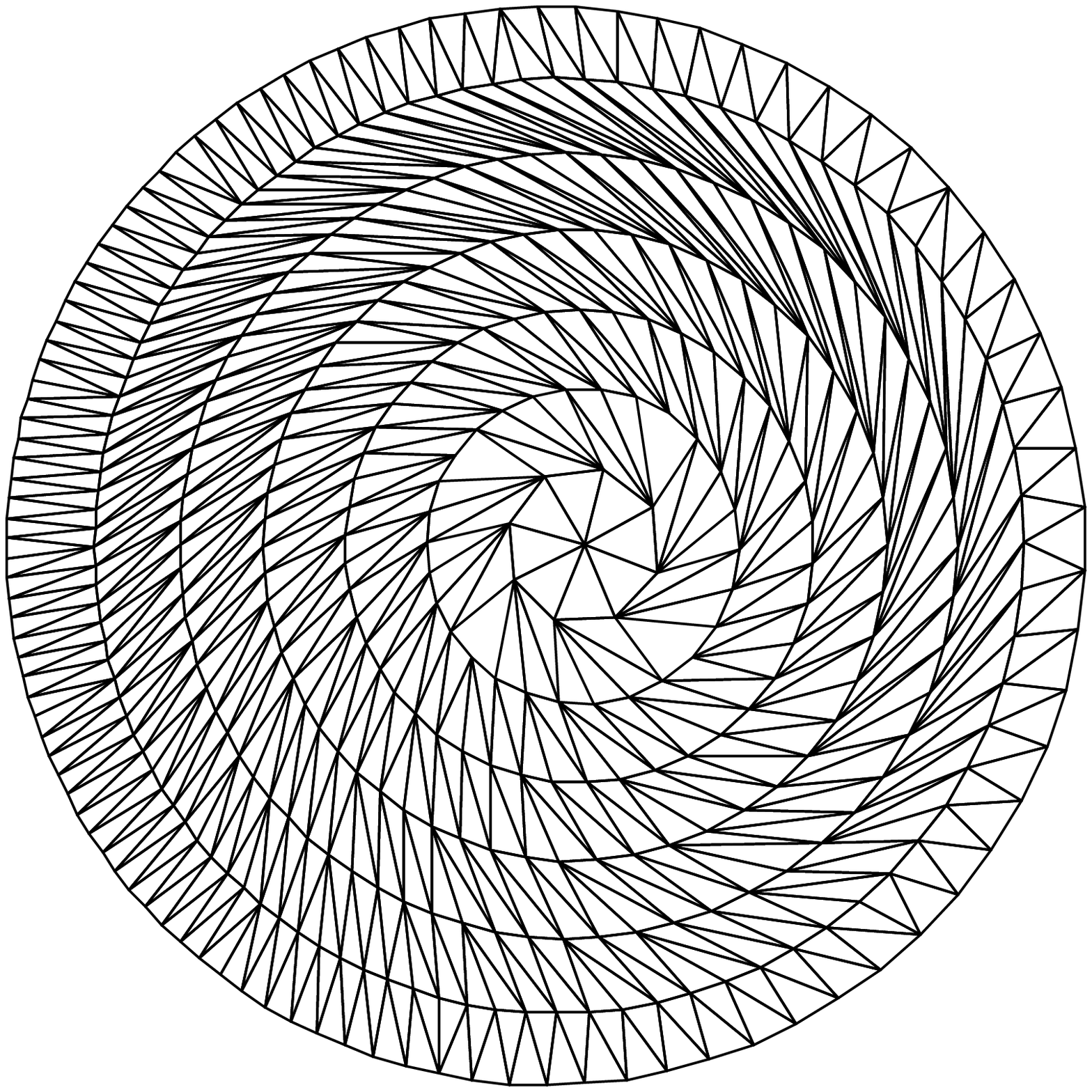}
\caption{Reduced resolution unstructured meshes at different toroidal angles, with node connectivity indicated, for a circular tokamak case. Note that the poloidal position of the nodes changes to accommodate the magnetic shear.}
\label{tok_meshes}
\end{figure}

\begin{figure}
\includegraphics[origin=c,scale=0.2]{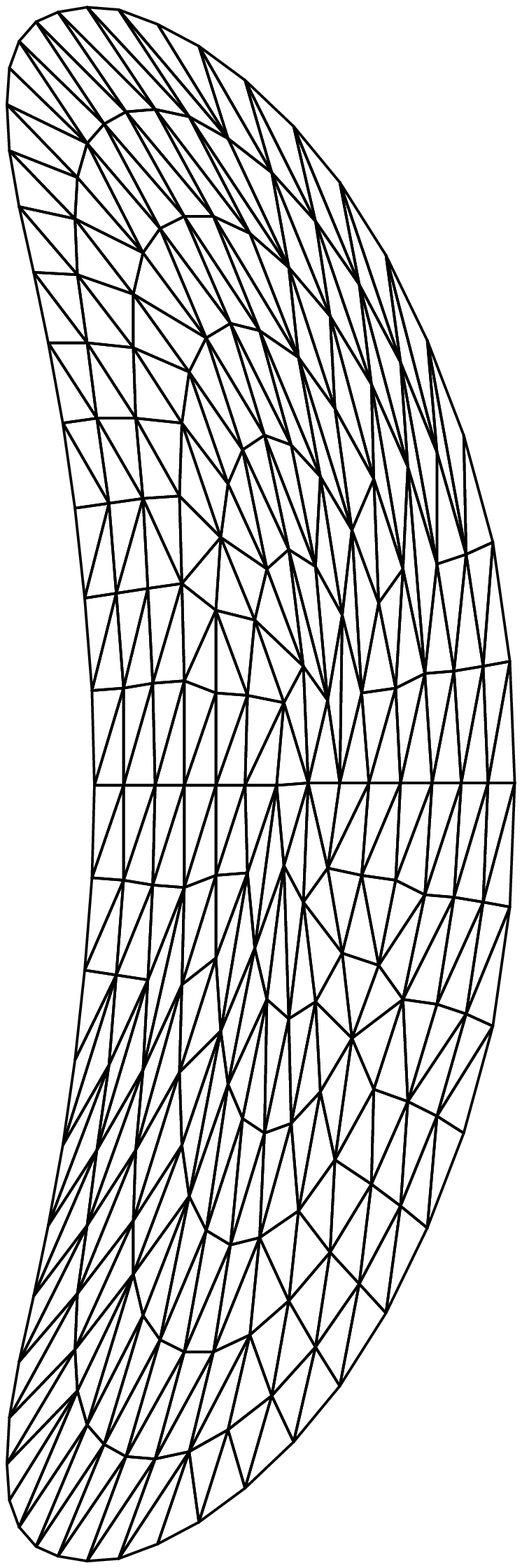}
\includegraphics[origin=c,scale=0.2]{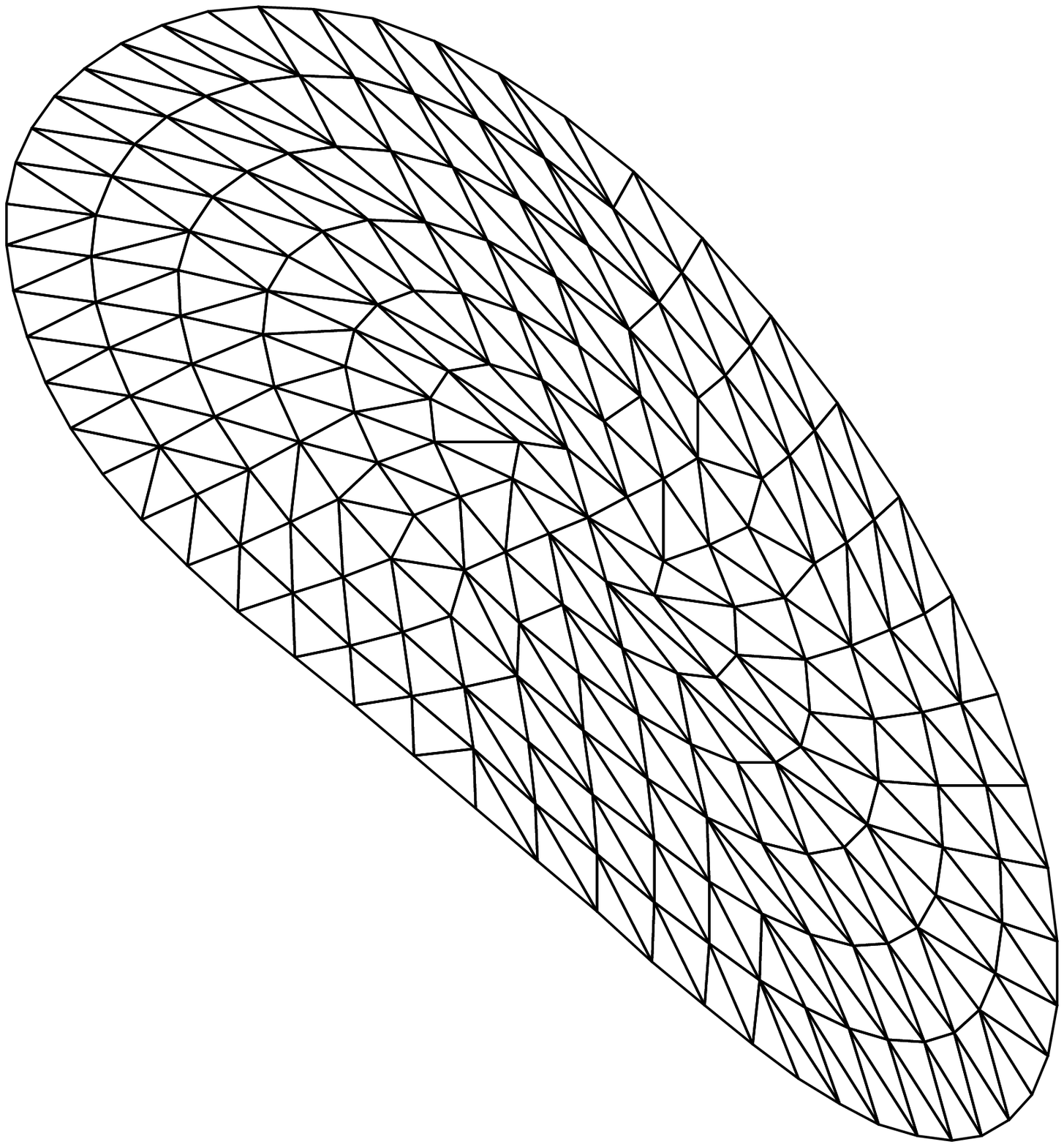}
\includegraphics[origin=c,scale=0.2]{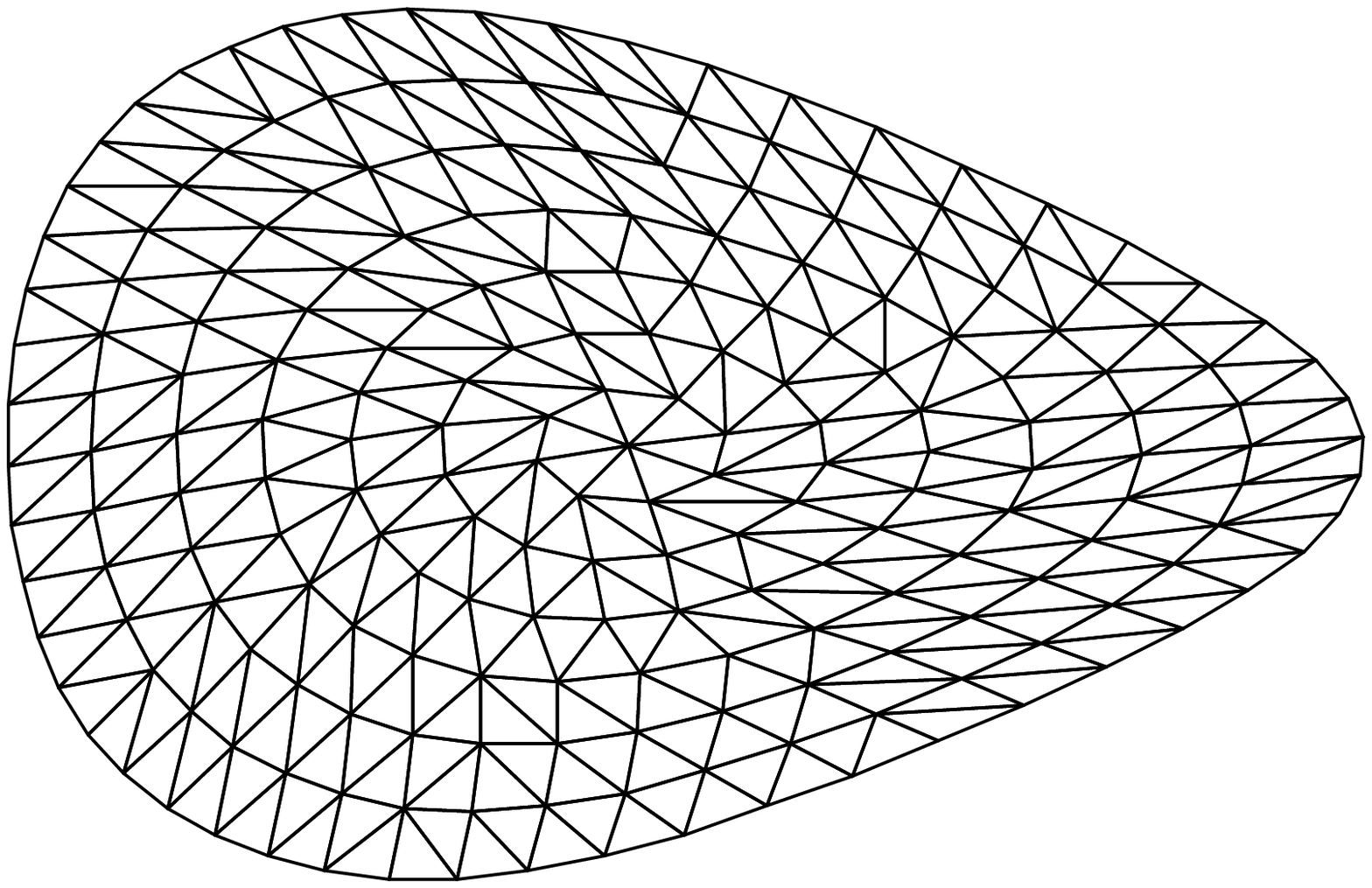}
\caption{Reduced resolution unstructured meshes for the Wendelstein 7-X stellarator at different toroidal positions within half a field period, illustrating the bean and triangular shaped cross-sections.}
\label{w7x_meshes}
\end{figure}

In XGC1, the Poisson equation is solved on a series of identical unstructured meshes composed of nodes placed on flux surfaces and mapped to positions on the cylindrical coordinate grid.
Nodes are placed such that, for the greatest practical proportion of each mesh, tracing a field line from a node on one plane to a neighbouring plane will arrive close to another node.
In XGC-S, unstructured meshes must have varying geometry at different toroidal locations.
A series of unstructured meshes is therefore produced based on VMEC output~\cite{moritaka_submitted} with nodes lying equidistantly on surfaces in PEST poloidal angle, and following field lines with toroidal angle.
In Figure~\ref{tok_meshes}, stylised (reduced radial and poloidal resolution) meshes produced for XGC-S are plotted at different toroidal angles to illustrate the twisting of nodes in toroidal angle.
Similar stylised meshes for Wendelstein 7-X geometry are shown in Figure~\ref{w7x_meshes}, illustrating also the change in mesh geometry with toroidal variation of the plasma boundary.

The use of unstructured meshes is a key distinguishing feature of the XGC code from the EUTERPE code, and other major global gyrokinetic codes.
It facilitates extension of the simulation domain into the edge region, where the magnetic field can be irregular.

The Poisson equation is currently solved independently on each planar mesh, with the electric field calculated in 3D using a finite differences method along the equilibrium magnetic field lines between the planes.
This is a valid approximation so long as the phenomena of interest are highly elongated along the magnetic field lines.

\subsection{Tokamak electrostatic ITG verification}

The 3D equilibria and mapping techniques for the stellarator code were previously verified~\cite{cole19}.
In that work, particle tracing studies were performed and collisionless orbit loss fractions compared when calculated with the EUTERPE, BEAMS3D~\cite{mcmillan14} and XGC codes.
They showed good agreement, and this capability was then used to investigate collisionless particle confinement in proposed stellarator reactors~\cite{moritaka_submitted,cole19}.

As an initial verification of the extensions to the stellarator code for solving the Poisson equation for ion-scale microinstabilities, we compare to a previously published, well-benchmarked calculation.
This calculation has been performed using the tokamak codes XGC1, ORB5 and global GENE, finding good agreement~\cite{merlo18}.
In this case, a circular tokamak geometry is chosen (Case V~\cite{burckel10}).
An electrostatic delta-$f$ gyrokinetic calculation to the last closed flux surface is performed, with analytic background profiles that localise a strong temperature gradient at the central flux surface.
The analytical form of the background plasma profiles is given in detail in reference~\cite{merlo18}, Section III.

Although the tokamak XGC1 and new XGC-S versions should, in this case, produce the same result, there are significant implementation differences.
XGC-S, for instance, uses a series of non-axisymmetric unstructured meshes to solve the Poisson equation, where XGC-1 solves the Poisson equation on the same axisymmetric mesh at each toroidal location~\cite{moritaka_submitted}.
Another is the interpolation of the equilibrium magnetic field, which in the extended stellarator case is performed using a 3D rather than 2D spline interpolation.
This equilibrium magnetic field model has been verified previously~\cite{moritaka_submitted,cole19}.
%XGC-S uses an optimised ion marker push kernel that is not present in XGC1.
%This development is not directly related to the geometry extension and does not alter the physics model.
%Its numerical properties and advantages will be described elsewhere~\cite{moritaka_prep}.

\begin{figure}
\includegraphics[origin=c,scale=0.6]{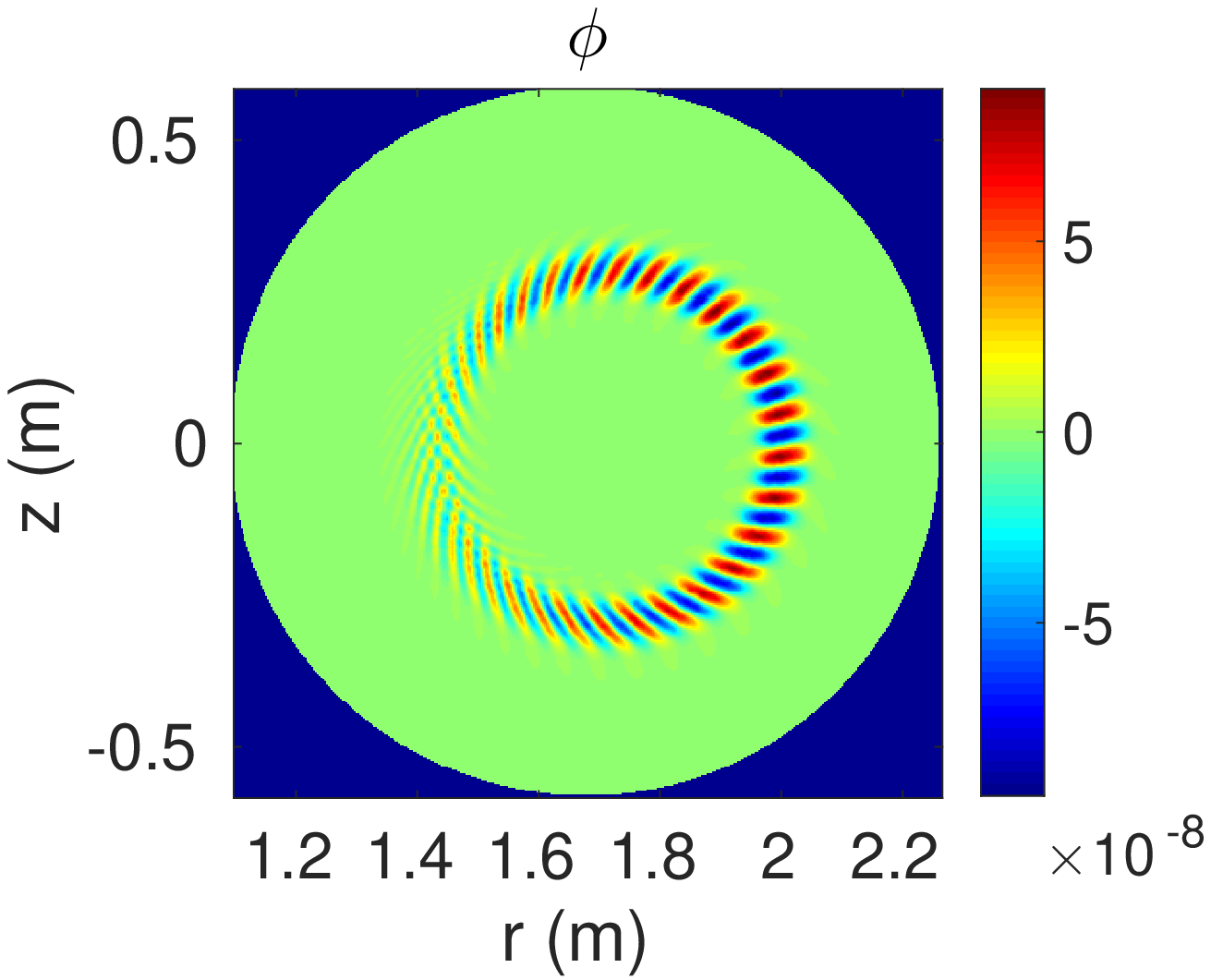}\includegraphics[origin=c,scale=0.6]{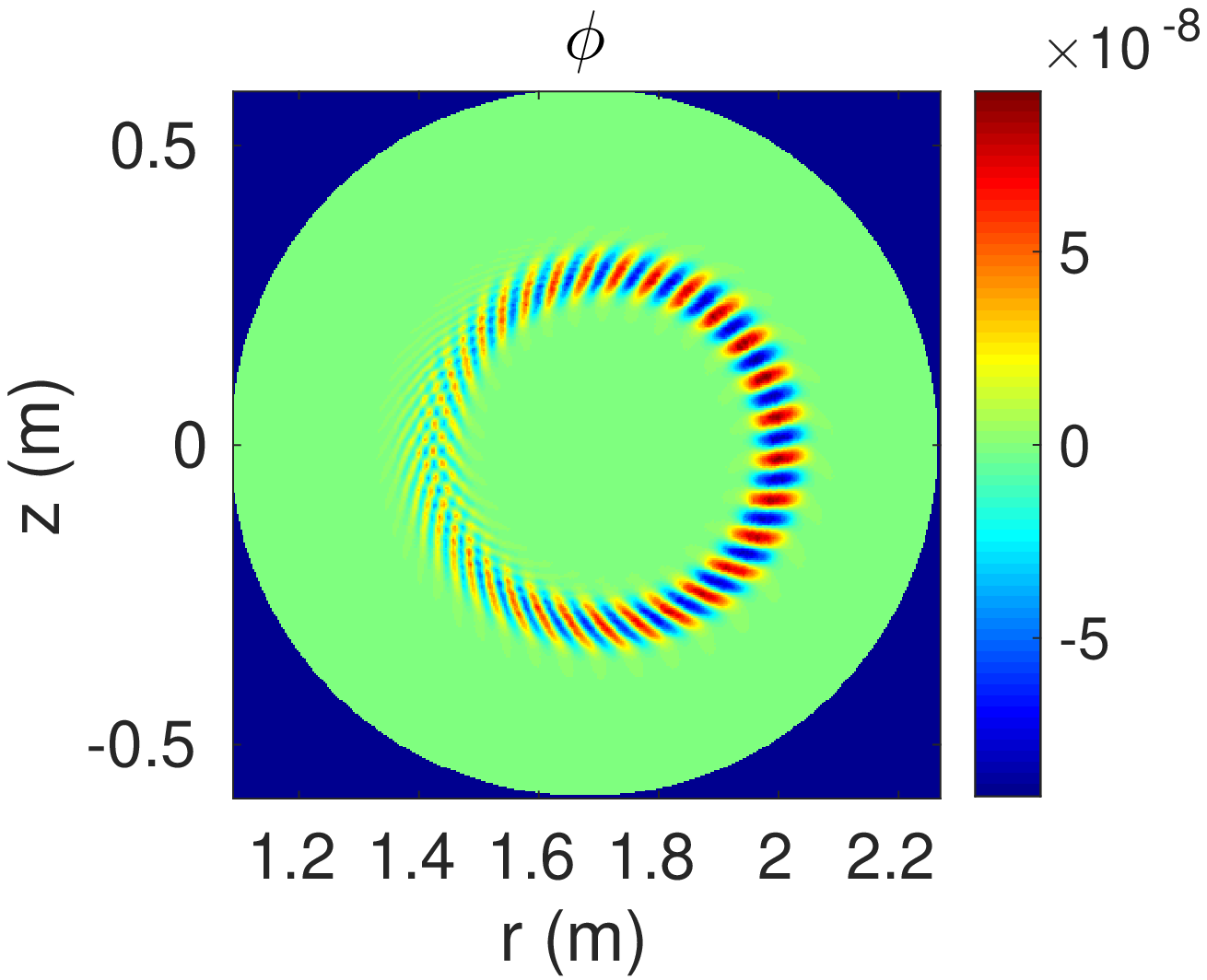}

\includegraphics[origin=c,scale=0.53]{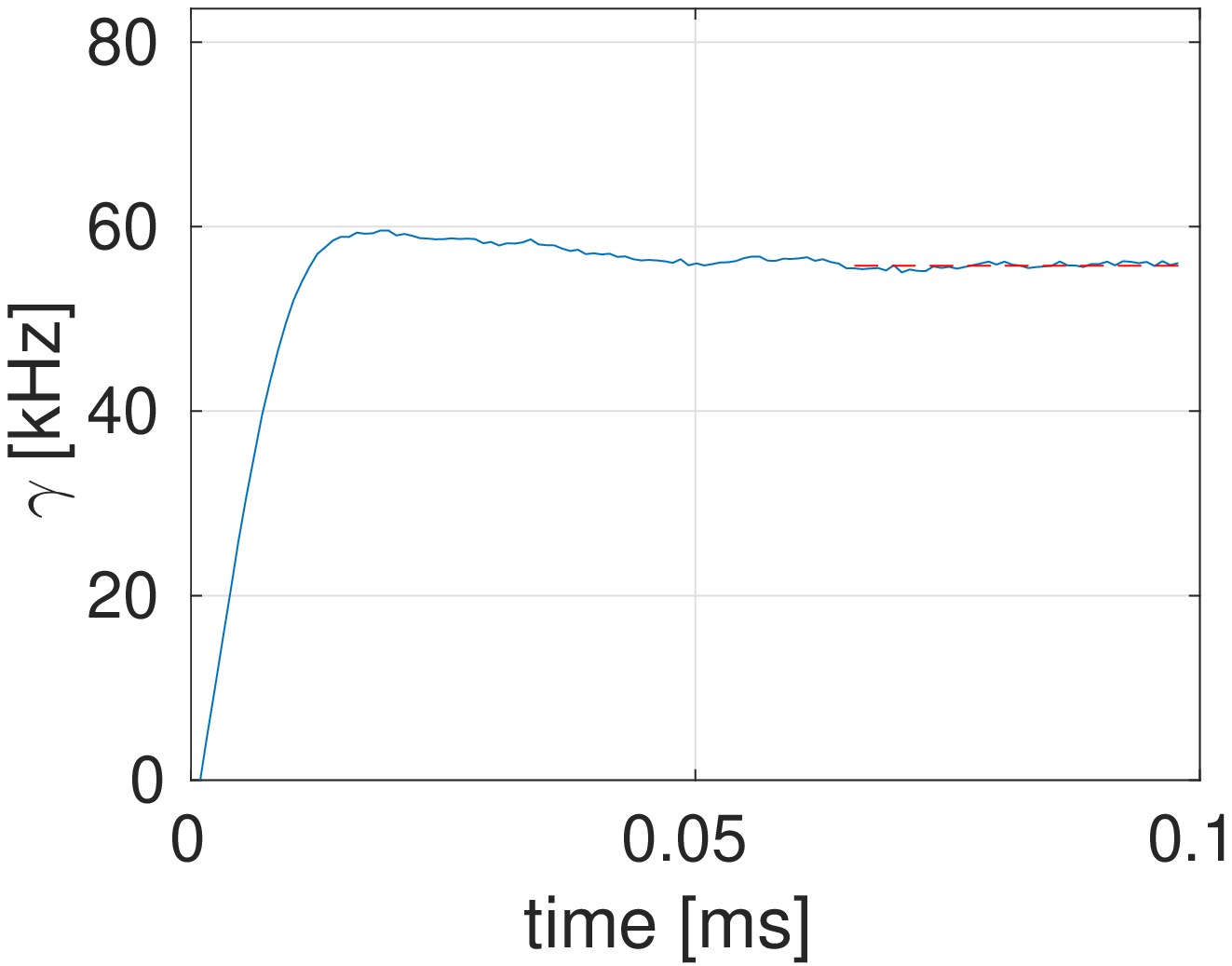}\includegraphics[origin=c,scale=0.53]{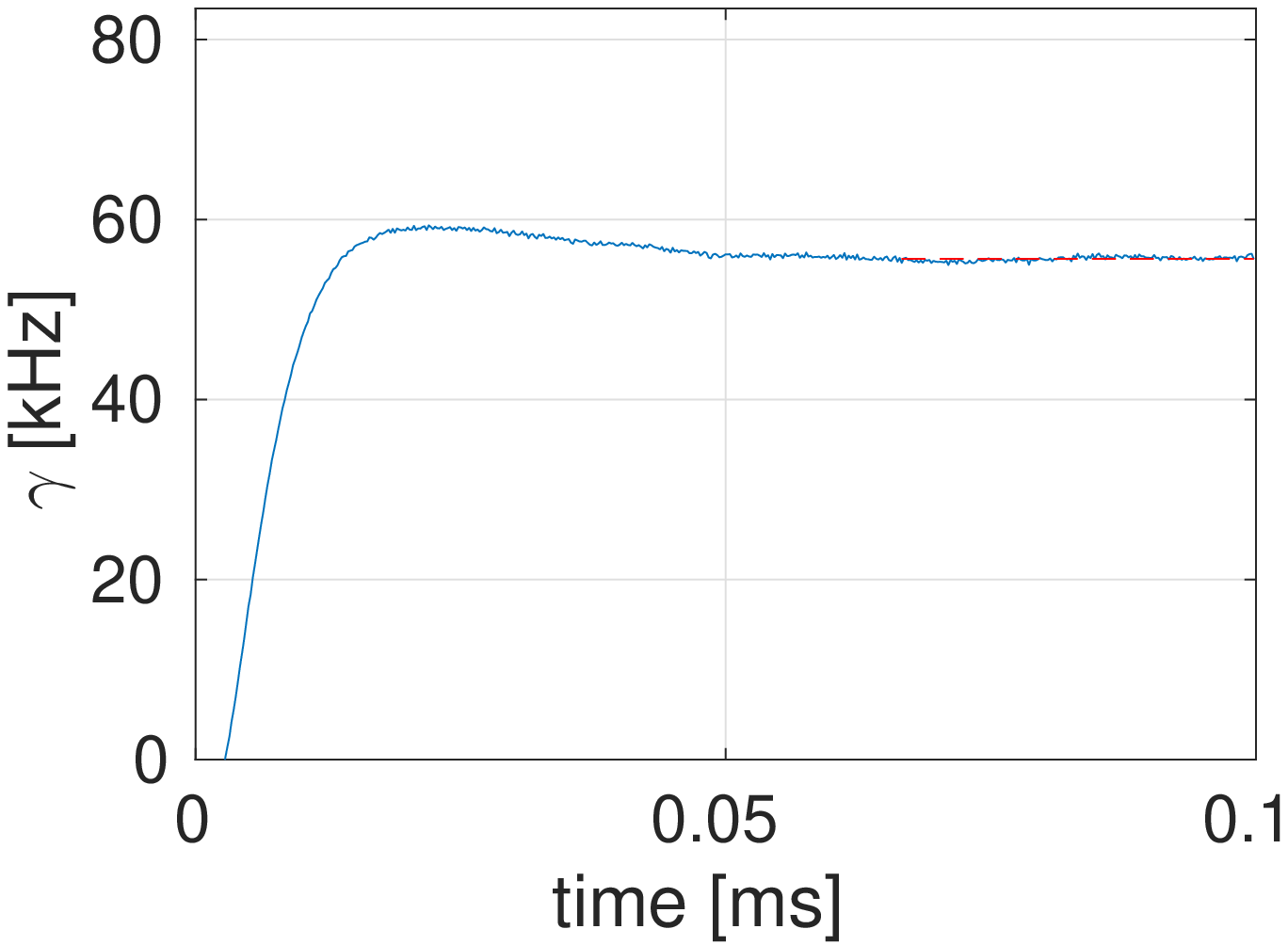}
\caption{\small left: tokamak version XGC1, right: stellarator version XGC-S}
\label{s_1_comp}
\end{figure}

In Figure~\ref{s_1_comp}, the calculated mode structure and growth rate are compared for this case using the two code versions.
In the stellarator version XGC-S the calculated growth rate is $\gamma_{XGC-S}= 55.63~\mathrm{kHz}$, while in the tokamak XGC1 version the calculated growth rate is $\gamma_{XGC1}= 55.75~\mathrm{kHz}$, giving a discrepancy of $0.22\%$.
Perfect agreement is not expected due to the difference in meshing; in particular, the region in which nodes are not field aligned (unavoidable on irrational surfaces) is necessarily not the same due to the difference in mesh design.
Nonetheless, the discrepancy is remarkably small, implying that the difference in meshing may not be important in calculating the ITG growth rate.
In both cases, the dominant toroidal mode number $n=24$ is chosen by simulating a $1/24$th toroidal wedge of the tokamak, and the dominant poloidal mode is correctly calculated to be $m=33$, as can be predicted considering the local value of the safety factor, $q=1.375$.
The number of toroidal slices (poloidal planes) within the wedge is $8$, as in the previous work, and the number of marker particles used is comparable within an order of magnitude ($10^7$).

\section{ITG microinstabilities in Wendelstein 7-X}
\label{w7x_results}

The only published result for simulation of ion-scale microinstabilities in stellarator geometry with a radial global model was obtained with the EUTERPE code, in comparison with flux tube ensemble (surface global) GENE simulations~\cite{helander15,riemann16}.
In this case, an artificial radially localised temperature gradient was used to approximate the local limit.
This case is chosen as a robust test for the model and implementation of XGC-S for linear electrostatic microinstabilities.

The magnetic geometry is provided by a fixed boundary VMEC run corresponding to the Wendelstein 7-X high mirror configuration.
The temperature gradient chosen has the piecewise form
\begin{equation}
\frac{\mathrm{d}~\mathrm{ln}~T}{\mathrm{d}~s} = -\sqrt{2}\left(\frac{1}{2}-\mid s-\frac{1}{2} \mid \right)\frac{a}{L_T},
\end{equation}
where $T$ is the ion temperature, $s$ is the normalised flux, $a$ is the minor radius (here, $0.505~\mathrm{m}$) and $L_T$ is the temperature gradient scale length.
A temperature gradient of this form localises the ITG drive around the central flux surface at $s=0.5$.
This allows a numerically clean simulation that can easily be compared with other codes, including surface global but radially local codes.
In order to maximise the ITG growth rate for this initial run, we choose not to apply a density gradient.
A density gradient is typically stabilising.
Instead, we choose $n_{i0} = n_{e0} = 3\times 10^{19}~\mathrm{m^{-3}}$ at all locations, a typical value for low density operation in W7-X experiments.

\begin{figure}
\includegraphics[origin=c,scale=0.5,angle=-90,origin=c]{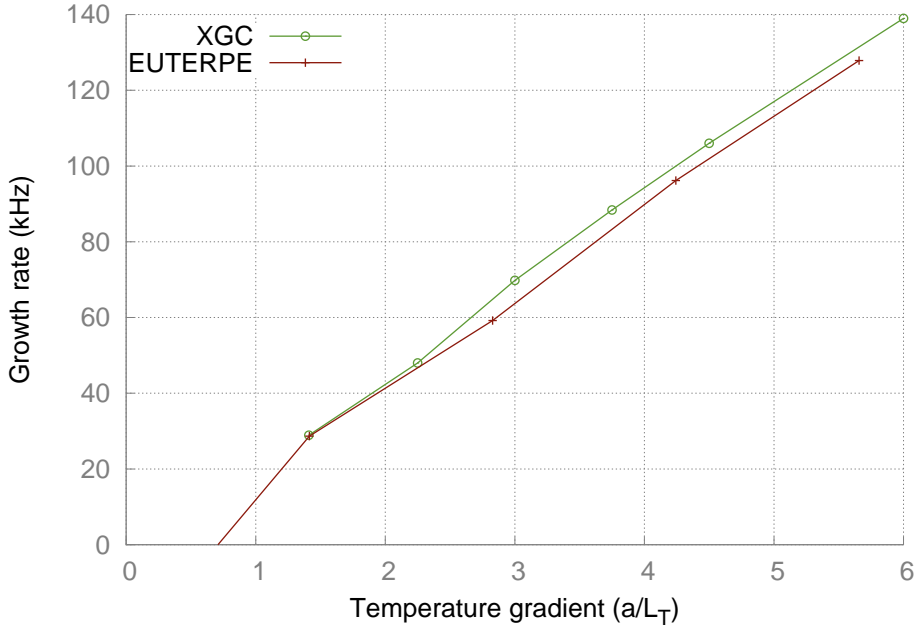}
\caption{\small Linear growth rates obtained for an ITG case in Wendelstein 7-X with the XGC and EUTERPE global gyrokinetic codes.}
\label{w7x_eut_xgc}
\end{figure}

With the EUTERPE code, this case has been run linearly.
We therefore perform an initial set of linear runs with the XGC-S code.
For the field calculation XGC-S is run with 256 toroidal domains per field period of W7-X, each mesh with approximately $1.4\times10^{5}$ mesh nodes.
As in the EUTERPE case, only one field period is simulated.
Mesh nodes lie on flux surfaces, chosen to be equidistant in $s$ with a radial resolution $N_s=64$ locations, as in the EUTERPE simulations.
The number of nodes per flux surface varies throughout a cross-section.
At $s=0.5$, there are approximately $6\times 10^3$ nodes distributed equidistantly in PEST poloidal angle.
In total, approximately $2.5\times 10^9$ numerical marker particles are used.
As in the published EUTERPE case, the long wavelength approximation is used in solving for the electrostatic potential.
The simulation domain radially is $0 < s < 0.95$.

\begin{figure}
\hspace{-4cm}\includegraphics[scale=0.3]{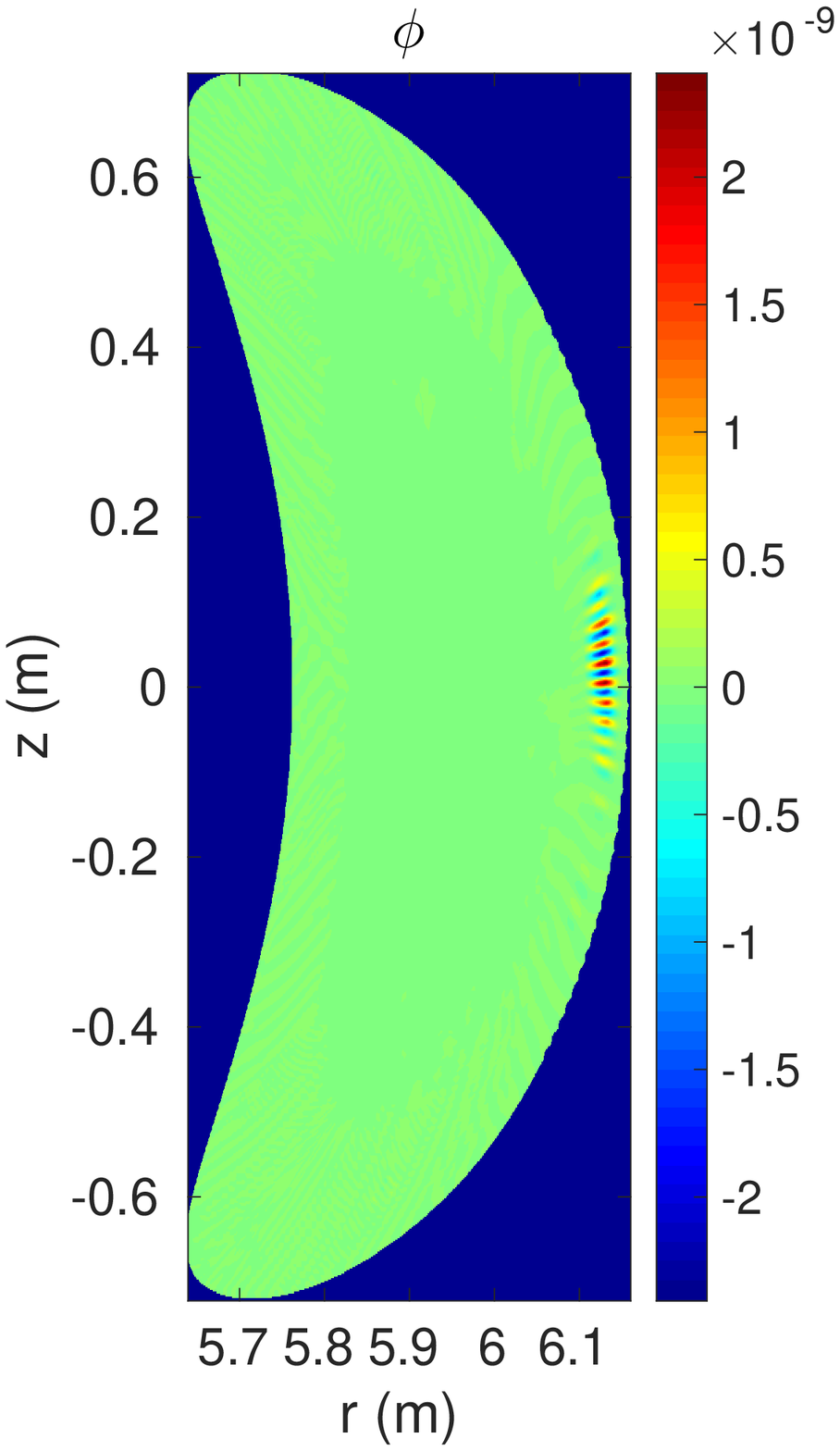}\includegraphics[scale=0.6]{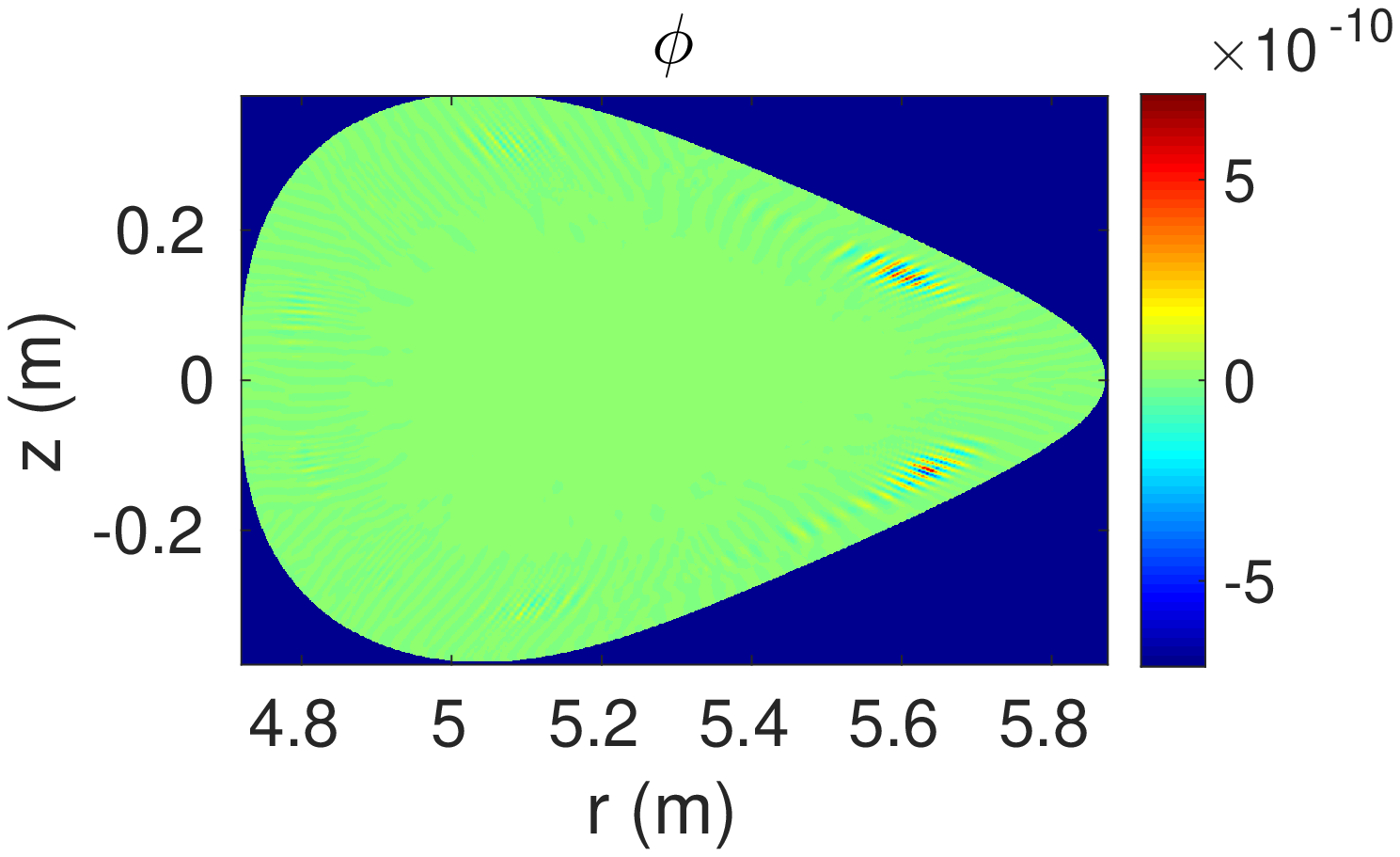}
\caption{Plots of the small linear perturbed electrostatic potential, $\phi$, calculated for $a/L_T=1.41$ at two toroidal positions within one period of W7-X. These correspond to the bean (left) and triangle-shaped (right) cross-sections.}
\label{w7x_dpot}
\end{figure}

In figure~\ref{w7x_eut_xgc}, the linear growth rate dependence on the temperature gradient strength parameter $a/L_T$ is compared between the XGC and EUTERPE codes~\cite{riemann16}.
Good qualitative and quantitative agreement is seen, with discrepancy in trend lines of less than $5\%$.
The mode structure in terms of the perturbed electrostatic potential is shown in Figure~\ref{w7x_dpot}, as calculated by XGC.
The potential is plotted at two poloidal cross-sections within the simulated field period of W7-X, at $\varphi=0$ and $\pi/5$, which correspond to the bean and triangle shaped cross-sections.
This case has a temperature gradient strength factor $a/L_T=1.41$.
Unlike in a tokamak, strong poloidal, and toroidal, localisation is expected and observed in the perturbation of fields and density.
This is seen with XGC.

\begin{figure}
\includegraphics[origin=c,scale=0.6]{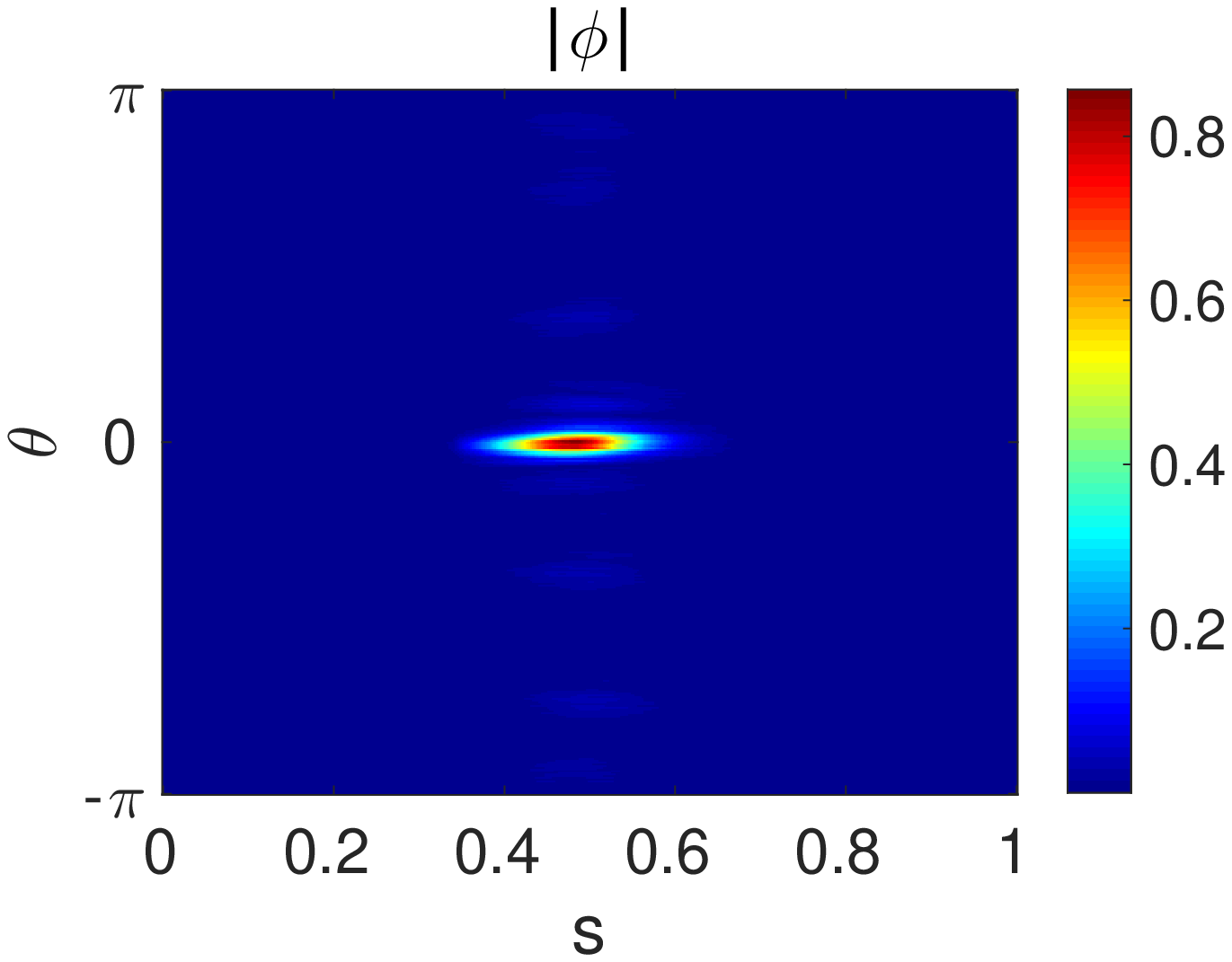}\includegraphics[origin=c,scale=0.6]{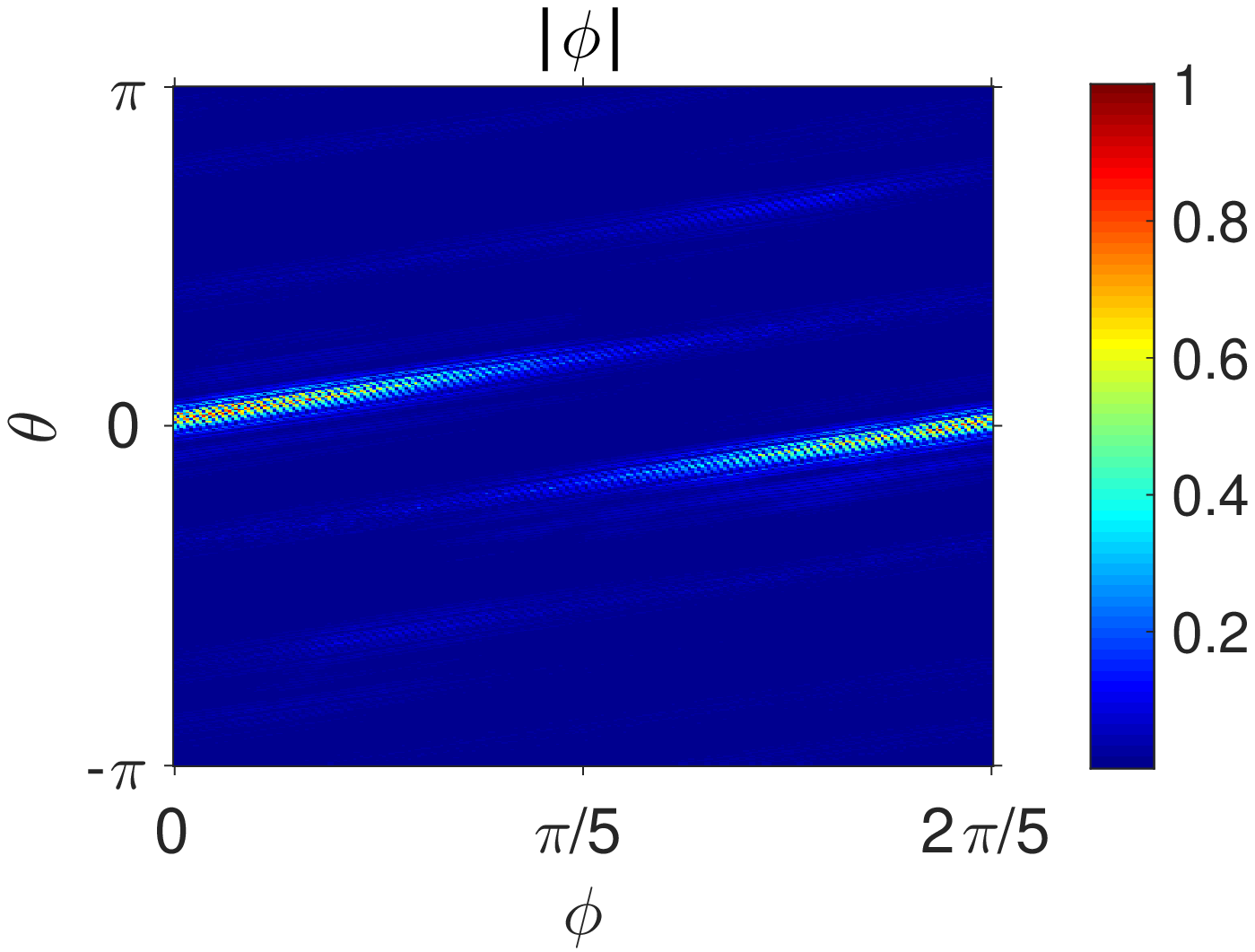}
\caption{\small Electrostatic potential structure plotted in terms of normalised poloidal flux and PEST poloidal angle at zero toroidal angle (left), and in terms of toroidal and PEST poloidal angle at $s=0.5$ (right). The colour scale is normalised to the maximum value in each plot.}
\label{xgc_1_41_sth}
\end{figure}

In Figure~\ref{xgc_1_41_sth}, the electrostatic potential structure is plotted in terms of $s$ and $\theta$ at zero toroidal angle (bean-shaped cross-section, left) and in terms of $\theta$ and $\phi$ at $s=0.5$ (right).
In both cases, $\theta$ is the PEST poloidal angle.
The left-hand plot can be qualitatively compared with Figure~4 in Ref.~\onlinecite{riemann16}.
However, exact comparison is not possible because here the real electrostatic potential is plotted without phase factor transformation, whereas in the reference the absolute value of a complex electrostatic potential is plotted with phase factor transformation.
The phase factor transformation obscures the fine scale structure by renormalising $n=0$ and $m=0$ to some finite value for the toroidal and poloidal mode numbers.
Nonetheless, the key feature of a perturbation localised close to the $s=0.5$ surface around $\theta=0$ is reproduced.

\begin{figure}
\includegraphics[origin=c,scale=0.6,origin=c]{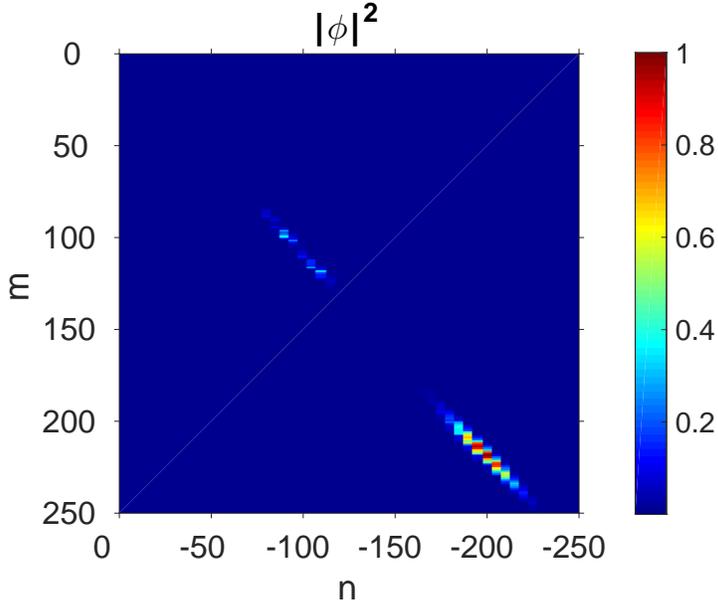}
\caption{The power spectrum of the linear ITG mode in terms of poloidal ($m$) and toroidal ($n$) mode numbers calculated by XGC-S for the $a/L_T=1.41$ case. Here $n$ is a full torus value, but limited to one mode family ($n = 0, 5, 10...$). $|\phi^2|$ is normalised to the maximum value.}
\label{w7x_mode_numbers}
\end{figure}

In Figure~\ref{w7x_mode_numbers}, the power spectrum is plotted: $|\phi|^2$ in terms of the toroidal and poloidal mode numbers, $n$ and $m$.
This can be compared to Figure~6 (left) in Ref.~\onlinecite{riemann16}.
Note that in the EUTERPE case, the phase factor transformation and Fourier filtering mean that only components of $\phi$ with values of $n$ and $m$ close to the expected solution are retained in the simulation.
In XGC-S, no phase factor transformation or filtering is used.
One can see that, as expected, the mode follows the line of resonance around $n+\iota m=0$.

\section{Conclusion}
\label{conclusion}

In this paper XGC-S, the stellarator version of the full volume global total-$f$ gyrokinetic code XGC, has been used to simulate, and verify the code for, linear ITG microinstabilities in a circular tokamak benchmark case, and a Wendelstein 7-X case.
XGC-S extends the existing XGC1 code~\cite{ku09,ku15,ku18} by the implementation of an interface for magnetic equilibria from the VMEC code and the ability to generate and utilise toroidally varying unstructured meshes for solving for the electrostatic potential with a toroidally varying plasma boundary and magnetic field~\cite{moritaka_submitted}.
This code has previously been benchmarked and applied for orbit tracing and neoclassical studies~\cite{cole19,moritaka_submitted}.

As an initial verification study, XGC-S has been used to reproduce a circular tokamak benchmark previously performed with the XGC1, ORB5 and global GENE codes~\cite{merlo18}.
Close agreement has been found between the two versions for ITG mode structure and growth rate.
The linear growth rates calculated by XGC1 and XGC-S differed by less than $1\%$.

Having established the validity of the code improvements by direct comparison to previous XGC results, the verification has been extended to stellarator geometry.
Linear simulations of ITG microinstabilities in the Wendelstein 7-X have been performed with a localised ion temperature gradient and flat density gradient, in comparison with published results from the EUTERPE code~\cite{helander15,riemann16}.
Linear growth rates and mode numbers have been directly compared, finding good agreement.
The calculated mode structure also shows good qualitative agreement.

In on-going work with the XGC-S code, neoclassical physics is being investigated in the Large Helical Device (LHD) heliotron.
In connection with this work the XGC-S physics model including field solver is being extended to the first wall~\cite{moritaka_submitted} and for a total-$f$ capability.
Nonlinear turbulence simulations are envisioned, opening up the possibility of comparison with experimental measurements.
In the outlook, XGC-S is an appropriate tool for modelling of neoclassical and turbulent physics simultaneously, and studying their nonlocal interaction.

\begin{acknowledgments}
We thank P.~Helander, J.~Geiger and A.~Dinklage for useful discussions and assistance.
We gratefully acknowledge S.~Seal, M.~Cianciosa, J.~Geiger and S.~Hirshman for their work developing, maintaining and running the VMEC code.
This manuscript has been authored by Princeton University under Contract Number DE-AC02-09CH11466.
The United States Government retains and the publisher, by accepting the article for publication, acknowledges that the United States Government retains a non-exclusive, paid-up, irrevocable, world-wide license to publish or reproduce the published form of this manuscript, or allow others to do so, for United States Government purposes.
This research used resources of the National Energy Research Scientific Computing Center (NERSC), a U.S. Department of Energy Office of Science User Facility operated under Contract No. DE-AC02-05CH11231.
\end{acknowledgments}

\bibliography{full}

\providecommand{\noopsort}[1]{}\providecommand{\singleletter}[1]{#1}%
\begin{thebibliography}{32}%
\makeatletter
\providecommand \@ifxundefined [1]{%
 \@ifx{#1\undefined}
}%
\providecommand \@ifnum [1]{%
 \ifnum #1\expandafter \@firstoftwo
 \else \expandafter \@secondoftwo
 \fi
}%
\providecommand \@ifx [1]{%
 \ifx #1\expandafter \@firstoftwo
 \else \expandafter \@secondoftwo
 \fi
}%
\providecommand \natexlab [1]{#1}%
\providecommand \enquote  [1]{``#1''}%
\providecommand \bibnamefont  [1]{#1}%
\providecommand \bibfnamefont [1]{#1}%
\providecommand \citenamefont [1]{#1}%
\providecommand \href@noop [0]{\@secondoftwo}%
\providecommand \href [0]{\begingroup \@sanitize@url \@href}%
\providecommand \@href[1]{\@@startlink{#1}\@@href}%
\providecommand \@@href[1]{\endgroup#1\@@endlink}%
\providecommand \@sanitize@url [0]{\catcode `\\12\catcode `\$12\catcode
  `\&12\catcode `\#12\catcode `\^12\catcode `\_12\catcode `\%12\relax}%
\providecommand \@@startlink[1]{}%
\providecommand \@@endlink[0]{}%
\providecommand \url  [0]{\begingroup\@sanitize@url \@url }%
\providecommand \@url [1]{\endgroup\@href {#1}{\urlprefix }}%
\providecommand \urlprefix  [0]{URL }%
\providecommand \Eprint [0]{\href }%
\providecommand \doibase [0]{http://dx.doi.org/}%
\providecommand \selectlanguage [0]{\@gobble}%
\providecommand \bibinfo  [0]{\@secondoftwo}%
\providecommand \bibfield  [0]{\@secondoftwo}%
\providecommand \translation [1]{[#1]}%
\providecommand \BibitemOpen [0]{}%
\providecommand \bibitemStop [0]{}%
\providecommand \bibitemNoStop [0]{.\EOS\space}%
\providecommand \EOS [0]{\spacefactor3000\relax}%
\providecommand \BibitemShut  [1]{\csname bibitem#1\endcsname}%
\let\auto@bib@innerbib\@empty
%</preamble>
\bibitem [{\citenamefont {Stroth}(1998)}]{stroth98}%
  \BibitemOpen
  \bibfield  {author} {\bibinfo {author} {\bibfnamefont {U.}~\bibnamefont
  {Stroth}},\ }\href@noop {} {\bibfield  {journal} {\bibinfo  {journal} {Plasma
  Phys. Control. Fusion}\ }\textbf {\bibinfo {volume} {40}},\ \bibinfo {pages}
  {9} (\bibinfo {year} {1998})}\BibitemShut {NoStop}%
\bibitem [{\citenamefont {Motojima}\ \emph {et~al.}(2003)\citenamefont
  {Motojima}, \citenamefont {Ohyabu}, \citenamefont {Komori}, \citenamefont
  {Kaneko}, \citenamefont {Yamada}, \citenamefont {Kawahata}, \citenamefont
  {Nakamura}, \citenamefont {Ida}, \citenamefont {Akiyama},\ and\ \citenamefont
  {Ashikawa}}]{motojima03}%
  \BibitemOpen
  \bibfield  {author} {\bibinfo {author} {\bibfnamefont {O.}~\bibnamefont
  {Motojima}}, \bibinfo {author} {\bibfnamefont {N.}~\bibnamefont {Ohyabu}},
  \bibinfo {author} {\bibfnamefont {A.}~\bibnamefont {Komori}}, \bibinfo
  {author} {\bibfnamefont {O.}~\bibnamefont {Kaneko}}, \bibinfo {author}
  {\bibfnamefont {H.}~\bibnamefont {Yamada}}, \bibinfo {author} {\bibfnamefont
  {K.}~\bibnamefont {Kawahata}}, \bibinfo {author} {\bibfnamefont
  {Y.}~\bibnamefont {Nakamura}}, \bibinfo {author} {\bibfnamefont
  {K.}~\bibnamefont {Ida}}, \bibinfo {author} {\bibfnamefont {T.}~\bibnamefont
  {Akiyama}}, \ and\ \bibinfo {author} {\bibfnamefont {N.}~\bibnamefont
  {Ashikawa}},\ }\href@noop {} {\bibfield  {journal} {\bibinfo  {journal}
  {Nucl. Fusion}\ }\textbf {\bibinfo {volume} {43}},\ \bibinfo {pages} {1674}
  (\bibinfo {year} {2003})}\BibitemShut {NoStop}%
\bibitem [{\citenamefont {Deng}\ \emph {et~al.}(2015)\citenamefont {Deng},
  \citenamefont {Brower}, \citenamefont {Anderson}, \citenamefont {Anderson},
  \citenamefont {Briesemeister},\ and\ \citenamefont {Likin}}]{deng15}%
  \BibitemOpen
  \bibfield  {author} {\bibinfo {author} {\bibfnamefont {C.~B.}\ \bibnamefont
  {Deng}}, \bibinfo {author} {\bibfnamefont {D.~L.}\ \bibnamefont {Brower}},
  \bibinfo {author} {\bibfnamefont {D.~T.}\ \bibnamefont {Anderson}}, \bibinfo
  {author} {\bibfnamefont {F.~S.~B.}\ \bibnamefont {Anderson}}, \bibinfo
  {author} {\bibfnamefont {A.}~\bibnamefont {Briesemeister}}, \ and\ \bibinfo
  {author} {\bibfnamefont {K.~M.}\ \bibnamefont {Likin}},\ }\href@noop {}
  {\bibfield  {journal} {\bibinfo  {journal} {Nucl. Fusion}\ }\textbf {\bibinfo
  {volume} {55}},\ \bibinfo {pages} {123003} (\bibinfo {year}
  {2015})}\BibitemShut {NoStop}%
\bibitem [{\citenamefont {N{\"u}hrenberg}\ and\ \citenamefont
  {Zille}(1986)}]{nuehrenberg86}%
  \BibitemOpen
  \bibfield  {author} {\bibinfo {author} {\bibfnamefont {J.}~\bibnamefont
  {N{\"u}hrenberg}}\ and\ \bibinfo {author} {\bibfnamefont {R.}~\bibnamefont
  {Zille}},\ }\href@noop {} {\bibfield  {journal} {\bibinfo  {journal} {Phys.
  Lett. A}\ }\textbf {\bibinfo {volume} {114A}},\ \bibinfo {pages} {3 129}
  (\bibinfo {year} {1986})}\BibitemShut {NoStop}%
\bibitem [{\citenamefont {Wolf}\ \emph {et~al.}(2017)\citenamefont {Wolf},
  \citenamefont {Ali}, \citenamefont {Alonso}, \citenamefont {Baldzuhn},
  \citenamefont {Beidler}, \citenamefont {Beurskens}, \citenamefont
  {Biedermann}, \citenamefont {Bosch}, \citenamefont {Bozhenkov}, \citenamefont
  {Brakel}, \citenamefont {Dinklage}, \citenamefont {Feng}, \citenamefont
  {Fuchert}, \citenamefont {Geiger}, \citenamefont {Grulke},\ and\
  \citenamefont {Helander}}]{wolf17}%
  \BibitemOpen
  \bibfield  {author} {\bibinfo {author} {\bibfnamefont {R.~C.}\ \bibnamefont
  {Wolf}}, \bibinfo {author} {\bibfnamefont {A.}~\bibnamefont {Ali}}, \bibinfo
  {author} {\bibfnamefont {A.}~\bibnamefont {Alonso}}, \bibinfo {author}
  {\bibfnamefont {J.}~\bibnamefont {Baldzuhn}}, \bibinfo {author}
  {\bibfnamefont {C.}~\bibnamefont {Beidler}}, \bibinfo {author} {\bibfnamefont
  {M.}~\bibnamefont {Beurskens}}, \bibinfo {author} {\bibfnamefont
  {C.}~\bibnamefont {Biedermann}}, \bibinfo {author} {\bibfnamefont {H.~S.}\
  \bibnamefont {Bosch}}, \bibinfo {author} {\bibfnamefont {S.}~\bibnamefont
  {Bozhenkov}}, \bibinfo {author} {\bibfnamefont {R.}~\bibnamefont {Brakel}},
  \bibinfo {author} {\bibfnamefont {A.}~\bibnamefont {Dinklage}}, \bibinfo
  {author} {\bibfnamefont {Y.}~\bibnamefont {Feng}}, \bibinfo {author}
  {\bibfnamefont {G.}~\bibnamefont {Fuchert}}, \bibinfo {author} {\bibfnamefont
  {J.}~\bibnamefont {Geiger}}, \bibinfo {author} {\bibfnamefont
  {O.}~\bibnamefont {Grulke}}, \ and\ \bibinfo {author} {\bibfnamefont
  {P.}~\bibnamefont {Helander}},\ }\href@noop {} {\bibfield  {journal}
  {\bibinfo  {journal} {Nucl. Fusion}\ }\textbf {\bibinfo {volume} {57}},\
  \bibinfo {pages} {102020} (\bibinfo {year} {2017})}\BibitemShut {NoStop}%
\bibitem [{\citenamefont {Klinger}\ \emph {et~al.}(2019)\citenamefont
  {Klinger}, \citenamefont {Andreeva}, \citenamefont {Bozhenkov}, \citenamefont
  {Brandt}, \citenamefont {Burhenn}, \citenamefont {Buttensch{\"o}n},
  \citenamefont {Fuchert}, \citenamefont {Geiger}, \citenamefont {Grulke},
  \citenamefont {Laqua}, \citenamefont {Pablant}, \citenamefont {Rahbarnia},
  \citenamefont {Stange}, \citenamefont {von Stechow}, \citenamefont {Tamura},
  \citenamefont {Thomsen}, \citenamefont {Wegner},\ and\ \citenamefont
  {Bussiahn}}]{klinger19}%
  \BibitemOpen
  \bibfield  {author} {\bibinfo {author} {\bibfnamefont {T.}~\bibnamefont
  {Klinger}}, \bibinfo {author} {\bibfnamefont {T.}~\bibnamefont {Andreeva}},
  \bibinfo {author} {\bibfnamefont {S.~A.}\ \bibnamefont {Bozhenkov}}, \bibinfo
  {author} {\bibfnamefont {C.}~\bibnamefont {Brandt}}, \bibinfo {author}
  {\bibfnamefont {R.}~\bibnamefont {Burhenn}}, \bibinfo {author} {\bibfnamefont
  {B.}~\bibnamefont {Buttensch{\"o}n}}, \bibinfo {author} {\bibfnamefont
  {G.}~\bibnamefont {Fuchert}}, \bibinfo {author} {\bibfnamefont
  {B.}~\bibnamefont {Geiger}}, \bibinfo {author} {\bibfnamefont
  {O.}~\bibnamefont {Grulke}}, \bibinfo {author} {\bibfnamefont {H.~P.}\
  \bibnamefont {Laqua}}, \bibinfo {author} {\bibfnamefont {N.~A.}\ \bibnamefont
  {Pablant}}, \bibinfo {author} {\bibfnamefont {K.}~\bibnamefont {Rahbarnia}},
  \bibinfo {author} {\bibfnamefont {T.}~\bibnamefont {Stange}}, \bibinfo
  {author} {\bibfnamefont {A.}~\bibnamefont {von Stechow}}, \bibinfo {author}
  {\bibfnamefont {N.}~\bibnamefont {Tamura}}, \bibinfo {author} {\bibfnamefont
  {H.}~\bibnamefont {Thomsen}}, \bibinfo {author} {\bibfnamefont
  {T.}~\bibnamefont {Wegner}}, \ and\ \bibinfo {author} {\bibfnamefont
  {R.}~\bibnamefont {Bussiahn}},\ }\href@noop {} {\bibfield  {journal}
  {\bibinfo  {journal} {Nucl. Fusion}\ } (\bibinfo {year} {2019})}\BibitemShut
  {NoStop}%
\bibitem [{\citenamefont {Proll}\ \emph {et~al.}(2012)\citenamefont {Proll},
  \citenamefont {Helander}, \citenamefont {Connor},\ and\ \citenamefont
  {Plunk}}]{proll12}%
  \BibitemOpen
  \bibfield  {author} {\bibinfo {author} {\bibfnamefont {J.~H.~E.}\
  \bibnamefont {Proll}}, \bibinfo {author} {\bibfnamefont {P.}~\bibnamefont
  {Helander}}, \bibinfo {author} {\bibfnamefont {J.~W.}\ \bibnamefont
  {Connor}}, \ and\ \bibinfo {author} {\bibfnamefont {G.~G.}\ \bibnamefont
  {Plunk}},\ }\href@noop {} {\bibfield  {journal} {\bibinfo  {journal} {Phys.
  Rev. Lett.}\ }\textbf {\bibinfo {volume} {108}},\ \bibinfo {pages} {245002}
  (\bibinfo {year} {2012})}\BibitemShut {NoStop}%
\bibitem [{\citenamefont {Xanthopoulos}\ \emph {et~al.}(2014)\citenamefont
  {Xanthopoulos}, \citenamefont {Mynick}, \citenamefont {Helander},
  \citenamefont {Turkin}, \citenamefont {Plunk}, \citenamefont {Jenko},
  \citenamefont {G{\"o}rler}, \citenamefont {Told}, \citenamefont {Bird},\ and\
  \citenamefont {Proll}}]{xanthopoulos14}%
  \BibitemOpen
  \bibfield  {author} {\bibinfo {author} {\bibfnamefont {P.}~\bibnamefont
  {Xanthopoulos}}, \bibinfo {author} {\bibfnamefont {H.~E.}\ \bibnamefont
  {Mynick}}, \bibinfo {author} {\bibfnamefont {P.}~\bibnamefont {Helander}},
  \bibinfo {author} {\bibfnamefont {Y.}~\bibnamefont {Turkin}}, \bibinfo
  {author} {\bibfnamefont {G.~G.}\ \bibnamefont {Plunk}}, \bibinfo {author}
  {\bibfnamefont {F.}~\bibnamefont {Jenko}}, \bibinfo {author} {\bibfnamefont
  {T.}~\bibnamefont {G{\"o}rler}}, \bibinfo {author} {\bibfnamefont
  {D.}~\bibnamefont {Told}}, \bibinfo {author} {\bibfnamefont {T.}~\bibnamefont
  {Bird}}, \ and\ \bibinfo {author} {\bibfnamefont {J.~H.~E.}\ \bibnamefont
  {Proll}},\ }\href@noop {} {\bibfield  {journal} {\bibinfo  {journal} {Phys.\
  Rev.\ Lett.}\ }\textbf {\bibinfo {volume} {113}},\ \bibinfo {pages} {155001}
  (\bibinfo {year} {2014})}\BibitemShut {NoStop}%
\bibitem [{\citenamefont {Plunk}\ \emph {et~al.}(2019)\citenamefont {Plunk},
  \citenamefont {Xanthopoulos}, \citenamefont {Weir}, \citenamefont
  {Bozhenkov}, \citenamefont {Dinklage}, \citenamefont {Fuchert}, \citenamefont
  {Geiger}, \citenamefont {Hirsch}, \citenamefont {Hoefel}, \citenamefont
  {Jakubowski}, \citenamefont {Langenberg}, \citenamefont {Pablant},
  \citenamefont {Pasch}, \citenamefont {Stange}, \citenamefont {Zhang},\ and\
  \citenamefont {the W7-X~Team}}]{plunk19}%
  \BibitemOpen
  \bibfield  {author} {\bibinfo {author} {\bibfnamefont {G.}~\bibnamefont
  {Plunk}}, \bibinfo {author} {\bibfnamefont {P.}~\bibnamefont {Xanthopoulos}},
  \bibinfo {author} {\bibfnamefont {G.}~\bibnamefont {Weir}}, \bibinfo {author}
  {\bibfnamefont {S.}~\bibnamefont {Bozhenkov}}, \bibinfo {author}
  {\bibfnamefont {A.}~\bibnamefont {Dinklage}}, \bibinfo {author}
  {\bibfnamefont {G.}~\bibnamefont {Fuchert}}, \bibinfo {author} {\bibfnamefont
  {J.}~\bibnamefont {Geiger}}, \bibinfo {author} {\bibfnamefont
  {M.}~\bibnamefont {Hirsch}}, \bibinfo {author} {\bibfnamefont
  {U.}~\bibnamefont {Hoefel}}, \bibinfo {author} {\bibfnamefont
  {M.}~\bibnamefont {Jakubowski}}, \bibinfo {author} {\bibfnamefont
  {A.}~\bibnamefont {Langenberg}}, \bibinfo {author} {\bibfnamefont
  {N.}~\bibnamefont {Pablant}}, \bibinfo {author} {\bibfnamefont
  {E.}~\bibnamefont {Pasch}}, \bibinfo {author} {\bibfnamefont
  {T.}~\bibnamefont {Stange}}, \bibinfo {author} {\bibfnamefont
  {D.}~\bibnamefont {Zhang}}, \ and\ \bibinfo {author} {\bibnamefont {the
  W7-X~Team}},\ }\href@noop {} {\bibfield  {journal} {\bibinfo  {journal}
  {Phys. Rev. Lett.}\ }\textbf {\bibinfo {volume} {122}},\ \bibinfo {pages}
  {035002} (\bibinfo {year} {2019})}\BibitemShut {NoStop}%
\bibitem [{\citenamefont {Brizard}\ and\ \citenamefont {Hahm}(2007)}]{hahm07}%
  \BibitemOpen
  \bibfield  {author} {\bibinfo {author} {\bibfnamefont {A.~J.}\ \bibnamefont
  {Brizard}}\ and\ \bibinfo {author} {\bibfnamefont {T.~S.}\ \bibnamefont
  {Hahm}},\ }\href@noop {} {\bibfield  {journal} {\bibinfo  {journal} {Rev.
  Mod. Phys.}\ }\textbf {\bibinfo {volume} {79}},\ \bibinfo {pages} {421}
  (\bibinfo {year} {2007})}\BibitemShut {NoStop}%
\bibitem [{\citenamefont {Proll}, \citenamefont {Xanthopoulos},\ and\
  \citenamefont {Helander}(2013)}]{proll13}%
  \BibitemOpen
  \bibfield  {author} {\bibinfo {author} {\bibfnamefont {J.~H.~E.}\
  \bibnamefont {Proll}}, \bibinfo {author} {\bibfnamefont {P.}~\bibnamefont
  {Xanthopoulos}}, \ and\ \bibinfo {author} {\bibfnamefont {P.}~\bibnamefont
  {Helander}},\ }\href@noop {} {\bibfield  {journal} {\bibinfo  {journal}
  {Phys. Plasmas}\ }\textbf {\bibinfo {volume} {20}},\ \bibinfo {pages}
  {122506} (\bibinfo {year} {2013})}\BibitemShut {NoStop}%
\bibitem [{\citenamefont {Xanthopoulos}\ \emph {et~al.}(2016)\citenamefont
  {Xanthopoulos}, \citenamefont {Plunk}, \citenamefont {Zocco},\ and\
  \citenamefont {Helander}}]{xanthopoulos16}%
  \BibitemOpen
  \bibfield  {author} {\bibinfo {author} {\bibfnamefont {P.}~\bibnamefont
  {Xanthopoulos}}, \bibinfo {author} {\bibfnamefont {G.}~\bibnamefont {Plunk}},
  \bibinfo {author} {\bibfnamefont {A.}~\bibnamefont {Zocco}}, \ and\ \bibinfo
  {author} {\bibfnamefont {P.}~\bibnamefont {Helander}},\ }\href@noop {}
  {\bibfield  {journal} {\bibinfo  {journal} {Phys. Rev. X}\ }\textbf {\bibinfo
  {volume} {6}},\ \bibinfo {pages} {021033} (\bibinfo {year}
  {2016})}\BibitemShut {NoStop}%
\bibitem [{\citenamefont {Helander}\ \emph {et~al.}(2015)\citenamefont
  {Helander}, \citenamefont {Bird}, \citenamefont {Kleiber}, \citenamefont
  {Plunk}, \citenamefont {Proll}, \citenamefont {Riemann},\ and\ \citenamefont
  {Xanthopoulos}}]{helander15}%
  \BibitemOpen
  \bibfield  {author} {\bibinfo {author} {\bibfnamefont {P.}~\bibnamefont
  {Helander}}, \bibinfo {author} {\bibfnamefont {T.}~\bibnamefont {Bird}},
  \bibinfo {author} {\bibfnamefont {R.}~\bibnamefont {Kleiber}}, \bibinfo
  {author} {\bibfnamefont {G.~G.}\ \bibnamefont {Plunk}}, \bibinfo {author}
  {\bibfnamefont {J.~H.~E.}\ \bibnamefont {Proll}}, \bibinfo {author}
  {\bibfnamefont {J.}~\bibnamefont {Riemann}}, \ and\ \bibinfo {author}
  {\bibfnamefont {P.}~\bibnamefont {Xanthopoulos}},\ }\href@noop {} {\bibfield
  {journal} {\bibinfo  {journal} {Nucl.\ Fusion}\ }\textbf {\bibinfo {volume}
  {55}},\ \bibinfo {pages} {053030} (\bibinfo {year} {2015})}\BibitemShut
  {NoStop}%
\bibitem [{\citenamefont {Riemann}, \citenamefont {Kleiber},\ and\
  \citenamefont {Borchardt}(2016)}]{riemann16}%
  \BibitemOpen
  \bibfield  {author} {\bibinfo {author} {\bibfnamefont {J.}~\bibnamefont
  {Riemann}}, \bibinfo {author} {\bibfnamefont {R.}~\bibnamefont {Kleiber}}, \
  and\ \bibinfo {author} {\bibfnamefont {M.}~\bibnamefont {Borchardt}},\
  }\href@noop {} {\bibfield  {journal} {\bibinfo  {journal} {Plasma Phys.\
  Control.\ Fusion}\ }\textbf {\bibinfo {volume} {58}},\ \bibinfo {pages}
  {074001} (\bibinfo {year} {2016})}\BibitemShut {NoStop}%
\bibitem [{\citenamefont {Kornilov}\ \emph {et~al.}(2004)\citenamefont
  {Kornilov}, \citenamefont {Kleiber}, \citenamefont {Hatzky}, \citenamefont
  {Villard},\ and\ \citenamefont {Jost}}]{kornilov04}%
  \BibitemOpen
  \bibfield  {author} {\bibinfo {author} {\bibfnamefont {V.}~\bibnamefont
  {Kornilov}}, \bibinfo {author} {\bibfnamefont {R.}~\bibnamefont {Kleiber}},
  \bibinfo {author} {\bibfnamefont {R.}~\bibnamefont {Hatzky}}, \bibinfo
  {author} {\bibfnamefont {L.}~\bibnamefont {Villard}}, \ and\ \bibinfo
  {author} {\bibfnamefont {G.}~\bibnamefont {Jost}},\ }\href@noop {} {\bibfield
   {journal} {\bibinfo  {journal} {Phys.\ Plasmas}\ }\textbf {\bibinfo {volume}
  {11}},\ \bibinfo {pages} {3196} (\bibinfo {year} {2004})}\BibitemShut
  {NoStop}%
\bibitem [{\citenamefont {Matsuoka}, \citenamefont {Idomura},\ and\
  \citenamefont {Satake}(2018)}]{matsuoka18}%
  \BibitemOpen
  \bibfield  {author} {\bibinfo {author} {\bibfnamefont {S.}~\bibnamefont
  {Matsuoka}}, \bibinfo {author} {\bibfnamefont {Y.}~\bibnamefont {Idomura}}, \
  and\ \bibinfo {author} {\bibfnamefont {S.}~\bibnamefont {Satake}},\
  }\href@noop {} {\bibfield  {journal} {\bibinfo  {journal} {Phys.\ Plasmas}\
  }\textbf {\bibinfo {volume} {25}},\ \bibinfo {pages} {022510} (\bibinfo
  {year} {2018})}\BibitemShut {NoStop}%
\bibitem [{\citenamefont {Spong}\ \emph {et~al.}(2017)\citenamefont {Spong},
  \citenamefont {Holod}, \citenamefont {Todo},\ and\ \citenamefont
  {Osakabe}}]{spong17}%
  \BibitemOpen
  \bibfield  {author} {\bibinfo {author} {\bibfnamefont {D.~A.}\ \bibnamefont
  {Spong}}, \bibinfo {author} {\bibfnamefont {I.}~\bibnamefont {Holod}},
  \bibinfo {author} {\bibfnamefont {Y.}~\bibnamefont {Todo}}, \ and\ \bibinfo
  {author} {\bibfnamefont {M.}~\bibnamefont {Osakabe}},\ }\href@noop {}
  {\bibfield  {journal} {\bibinfo  {journal} {Nucl.\ Fusion}\ }\textbf
  {\bibinfo {volume} {57}},\ \bibinfo {pages} {086018} (\bibinfo {year}
  {2017})}\BibitemShut {NoStop}%
\bibitem [{\citenamefont {Todo}\ \emph {et~al.}(2017)\citenamefont {Todo},
  \citenamefont {Seki}, \citenamefont {Spong}, \citenamefont {Wang},
  \citenamefont {Suzuki}, \citenamefont {Yamamoto}, \citenamefont {Nakajima},\
  and\ \citenamefont {Osakabe}}]{todo17}%
  \BibitemOpen
  \bibfield  {author} {\bibinfo {author} {\bibfnamefont {Y.}~\bibnamefont
  {Todo}}, \bibinfo {author} {\bibfnamefont {R.}~\bibnamefont {Seki}}, \bibinfo
  {author} {\bibfnamefont {D.~A.}\ \bibnamefont {Spong}}, \bibinfo {author}
  {\bibfnamefont {H.}~\bibnamefont {Wang}}, \bibinfo {author} {\bibfnamefont
  {Y.}~\bibnamefont {Suzuki}}, \bibinfo {author} {\bibfnamefont
  {S.}~\bibnamefont {Yamamoto}}, \bibinfo {author} {\bibfnamefont
  {N.}~\bibnamefont {Nakajima}}, \ and\ \bibinfo {author} {\bibfnamefont
  {M.}~\bibnamefont {Osakabe}},\ }\href@noop {} {\bibfield  {journal} {\bibinfo
   {journal} {Phys. Plasmas}\ }\textbf {\bibinfo {volume} {24}},\ \bibinfo
  {pages} {081203} (\bibinfo {year} {2017})}\BibitemShut {NoStop}%
\bibitem [{\citenamefont {Ku}, \citenamefont {Chang},\ and\ \citenamefont
  {Diamond}(2009)}]{ku09}%
  \BibitemOpen
  \bibfield  {author} {\bibinfo {author} {\bibfnamefont {S.~H.}\ \bibnamefont
  {Ku}}, \bibinfo {author} {\bibfnamefont {C.~S.}\ \bibnamefont {Chang}}, \
  and\ \bibinfo {author} {\bibfnamefont {P.}~\bibnamefont {Diamond}},\
  }\href@noop {} {\bibfield  {journal} {\bibinfo  {journal} {Nucl.\ Fusion}\
  }\textbf {\bibinfo {volume} {49}},\ \bibinfo {pages} {115021} (\bibinfo
  {year} {2009})}\BibitemShut {NoStop}%
\bibitem [{\citenamefont {Ku}\ \emph {et~al.}(2015)\citenamefont {Ku},
  \citenamefont {Hager}, \citenamefont {Chang}, \citenamefont {Kwon},\ and\
  \citenamefont {Parker}}]{ku15}%
  \BibitemOpen
  \bibfield  {author} {\bibinfo {author} {\bibfnamefont {S.~H.}\ \bibnamefont
  {Ku}}, \bibinfo {author} {\bibfnamefont {R.}~\bibnamefont {Hager}}, \bibinfo
  {author} {\bibfnamefont {C.~S.}\ \bibnamefont {Chang}}, \bibinfo {author}
  {\bibfnamefont {J.}~\bibnamefont {Kwon}}, \ and\ \bibinfo {author}
  {\bibfnamefont {S.}~\bibnamefont {Parker}},\ }\href@noop {} {\bibfield
  {journal} {\bibinfo  {journal} {J.\ Comp.\ Physics}\ }\textbf {\bibinfo
  {volume} {315}},\ \bibinfo {pages} {467} (\bibinfo {year}
  {2015})}\BibitemShut {NoStop}%
\bibitem [{\citenamefont {Ku}\ \emph {et~al.}(2018)\citenamefont {Ku},
  \citenamefont {Chang}, \citenamefont {Hager}, \citenamefont {Churchill},
  \citenamefont {Tynan}, \citenamefont {Cziegler}, \citenamefont {Greenwald},
  \citenamefont {Hughes}, \citenamefont {Parker}, \citenamefont {Adams},
  \citenamefont {D'Azevedo},\ and\ \citenamefont {Worley}}]{ku18}%
  \BibitemOpen
  \bibfield  {author} {\bibinfo {author} {\bibfnamefont {S.}~\bibnamefont
  {Ku}}, \bibinfo {author} {\bibfnamefont {C.~S.}\ \bibnamefont {Chang}},
  \bibinfo {author} {\bibfnamefont {R.}~\bibnamefont {Hager}}, \bibinfo
  {author} {\bibfnamefont {R.~M.}\ \bibnamefont {Churchill}}, \bibinfo {author}
  {\bibfnamefont {G.~R.}\ \bibnamefont {Tynan}}, \bibinfo {author}
  {\bibfnamefont {I.}~\bibnamefont {Cziegler}}, \bibinfo {author}
  {\bibfnamefont {M.}~\bibnamefont {Greenwald}}, \bibinfo {author}
  {\bibfnamefont {J.}~\bibnamefont {Hughes}}, \bibinfo {author} {\bibfnamefont
  {S.~E.}\ \bibnamefont {Parker}}, \bibinfo {author} {\bibfnamefont {M.~F.}\
  \bibnamefont {Adams}}, \bibinfo {author} {\bibfnamefont {E.}~\bibnamefont
  {D'Azevedo}}, \ and\ \bibinfo {author} {\bibfnamefont {P.}~\bibnamefont
  {Worley}},\ }\href@noop {} {\bibfield  {journal} {\bibinfo  {journal} {Phys.\
  Plasmas}\ }\textbf {\bibinfo {volume} {25}},\ \bibinfo {pages} {056107}
  (\bibinfo {year} {2018})}\BibitemShut {NoStop}%
\bibitem [{\citenamefont {Churchill}\ \emph {et~al.}(2017)\citenamefont
  {Churchill}, \citenamefont {Chang}, \citenamefont {Ku},\ and\ \citenamefont
  {Dominski}}]{churchill17}%
  \BibitemOpen
  \bibfield  {author} {\bibinfo {author} {\bibfnamefont {R.~M.}\ \bibnamefont
  {Churchill}}, \bibinfo {author} {\bibfnamefont {C.~S.}\ \bibnamefont
  {Chang}}, \bibinfo {author} {\bibfnamefont {S.}~\bibnamefont {Ku}}, \ and\
  \bibinfo {author} {\bibfnamefont {J.}~\bibnamefont {Dominski}},\ }\href@noop
  {} {\bibfield  {journal} {\bibinfo  {journal} {Plasma Phys.\ Control.\
  Fusion}\ }\textbf {\bibinfo {volume} {59}},\ \bibinfo {pages} {105014}
  (\bibinfo {year} {2017})}\BibitemShut {NoStop}%
\bibitem [{\citenamefont {Chang}\ \emph {et~al.}(2017)\citenamefont {Chang},
  \citenamefont {Ku}, \citenamefont {Tynan}, \citenamefont {Hager},
  \citenamefont {Churchill}, \citenamefont {Cziegler}, \citenamefont
  {Greenwald}, \citenamefont {Hubbard},\ and\ \citenamefont
  {Hughes}}]{chang17}%
  \BibitemOpen
  \bibfield  {author} {\bibinfo {author} {\bibfnamefont {C.}~\bibnamefont
  {Chang}}, \bibinfo {author} {\bibfnamefont {S.}~\bibnamefont {Ku}}, \bibinfo
  {author} {\bibfnamefont {G.}~\bibnamefont {Tynan}}, \bibinfo {author}
  {\bibfnamefont {R.}~\bibnamefont {Hager}}, \bibinfo {author} {\bibfnamefont
  {R.}~\bibnamefont {Churchill}}, \bibinfo {author} {\bibfnamefont
  {I.}~\bibnamefont {Cziegler}}, \bibinfo {author} {\bibfnamefont
  {M.}~\bibnamefont {Greenwald}}, \bibinfo {author} {\bibfnamefont
  {A.}~\bibnamefont {Hubbard}}, \ and\ \bibinfo {author} {\bibfnamefont
  {J.}~\bibnamefont {Hughes}},\ }\href@noop {} {\bibfield  {journal} {\bibinfo
  {journal} {Phys. Rev. Lett.}\ }\textbf {\bibinfo {volume} {118}},\ \bibinfo
  {pages} {175001} (\bibinfo {year} {2017})}\BibitemShut {NoStop}%
\bibitem [{\citenamefont {Moritaka}\ and\ \citenamefont
  {et~al.}(pted)}]{moritaka_submitted}%
  \BibitemOpen
  \bibfield  {author} {\bibinfo {author} {\bibfnamefont {T.}~\bibnamefont
  {Moritaka}}\ and\ \bibinfo {author} {\bibnamefont {et~al.}},\ }\href@noop {}
  {\bibfield  {journal} {\bibinfo  {journal} {Plasma}\ } (\bibinfo {year}
  {Accepted})}\BibitemShut {NoStop}%
\bibitem [{\citenamefont {Cole}\ \emph {et~al.}(2019)\citenamefont {Cole},
  \citenamefont {Hager}, \citenamefont {Moritaka}, \citenamefont {Lazerson},
  \citenamefont {Kleiber}, \citenamefont {Ku},\ and\ \citenamefont
  {Chang}}]{cole19}%
  \BibitemOpen
  \bibfield  {author} {\bibinfo {author} {\bibfnamefont {M.~D.~J.}\
  \bibnamefont {Cole}}, \bibinfo {author} {\bibfnamefont {R.}~\bibnamefont
  {Hager}}, \bibinfo {author} {\bibfnamefont {T.}~\bibnamefont {Moritaka}},
  \bibinfo {author} {\bibfnamefont {S.}~\bibnamefont {Lazerson}}, \bibinfo
  {author} {\bibfnamefont {R.}~\bibnamefont {Kleiber}}, \bibinfo {author}
  {\bibfnamefont {S.}~\bibnamefont {Ku}}, \ and\ \bibinfo {author}
  {\bibfnamefont {C.~S.}\ \bibnamefont {Chang}},\ }\href@noop {} {\bibfield
  {journal} {\bibinfo  {journal} {Phys. Plasmas}\ }\textbf {\bibinfo {volume}
  {26}},\ \bibinfo {pages} {032506} (\bibinfo {year} {2019})}\BibitemShut
  {NoStop}%
\bibitem [{\citenamefont {Hager}\ \emph {et~al.}(2016)\citenamefont {Hager},
  \citenamefont {Yoon}, \citenamefont {Ku}, \citenamefont {D'Azevedo},
  \citenamefont {Worley},\ and\ \citenamefont {Chang}}]{hager16}%
  \BibitemOpen
  \bibfield  {author} {\bibinfo {author} {\bibfnamefont {R.}~\bibnamefont
  {Hager}}, \bibinfo {author} {\bibfnamefont {E.~S.}\ \bibnamefont {Yoon}},
  \bibinfo {author} {\bibfnamefont {S.}~\bibnamefont {Ku}}, \bibinfo {author}
  {\bibfnamefont {E.~F.}\ \bibnamefont {D'Azevedo}}, \bibinfo {author}
  {\bibfnamefont {P.~H.}\ \bibnamefont {Worley}}, \ and\ \bibinfo {author}
  {\bibfnamefont {C.~S.}\ \bibnamefont {Chang}},\ }\href@noop {} {\bibfield
  {journal} {\bibinfo  {journal} {J.\ Comp.\ Physics}\ }\textbf {\bibinfo
  {volume} {315}},\ \bibinfo {pages} {644} (\bibinfo {year}
  {2016})}\BibitemShut {NoStop}%
\bibitem [{\citenamefont {Stotler}\ \emph {et~al.}(2017)\citenamefont
  {Stotler}, \citenamefont {Lang}, \citenamefont {Chang}, \citenamefont
  {Churchill},\ and\ \citenamefont {Ku}}]{stotler17}%
  \BibitemOpen
  \bibfield  {author} {\bibinfo {author} {\bibfnamefont {D.~P.}\ \bibnamefont
  {Stotler}}, \bibinfo {author} {\bibfnamefont {J.}~\bibnamefont {Lang}},
  \bibinfo {author} {\bibfnamefont {C.~S.}\ \bibnamefont {Chang}}, \bibinfo
  {author} {\bibfnamefont {R.~M.}\ \bibnamefont {Churchill}}, \ and\ \bibinfo
  {author} {\bibfnamefont {S.}~\bibnamefont {Ku}},\ }\href@noop {} {\bibfield
  {journal} {\bibinfo  {journal} {Nucl. Fusion}\ }\textbf {\bibinfo {volume}
  {57}},\ \bibinfo {pages} {8} (\bibinfo {year} {2017})}\BibitemShut {NoStop}%
\bibitem [{\citenamefont {Hirshman}\ and\ \citenamefont
  {Whitson}(1983)}]{hirshman83}%
  \BibitemOpen
  \bibfield  {author} {\bibinfo {author} {\bibfnamefont {S.~P.}\ \bibnamefont
  {Hirshman}}\ and\ \bibinfo {author} {\bibfnamefont {J.~C.}\ \bibnamefont
  {Whitson}},\ }\href@noop {} {\bibfield  {journal} {\bibinfo  {journal}
  {Phys.\ Fluids}\ }\textbf {\bibinfo {volume} {26}},\ \bibinfo {pages}
  {3553–68} (\bibinfo {year} {1983})}\BibitemShut {NoStop}%
\bibitem [{\citenamefont {McMillan}\ and\ \citenamefont
  {Lazerson}(2014)}]{mcmillan14}%
  \BibitemOpen
  \bibfield  {author} {\bibinfo {author} {\bibfnamefont {M.}~\bibnamefont
  {McMillan}}\ and\ \bibinfo {author} {\bibfnamefont {S.~A.}\ \bibnamefont
  {Lazerson}},\ }\href@noop {} {\bibfield  {journal} {\bibinfo  {journal}
  {Plasma Phys.\ Control.\ Fusion}\ }\textbf {\bibinfo {volume} {56}},\
  \bibinfo {pages} {095019} (\bibinfo {year} {2014})}\BibitemShut {NoStop}%
\bibitem [{\citenamefont {Merlo}\ \emph {et~al.}(2018)\citenamefont {Merlo},
  \citenamefont {Dominski}, \citenamefont {Bhattacharjee}, \citenamefont
  {Chang}, \citenamefont {Jenko}, \citenamefont {Ku}, \citenamefont {Lanti},\
  and\ \citenamefont {Parker}}]{merlo18}%
  \BibitemOpen
  \bibfield  {author} {\bibinfo {author} {\bibfnamefont {G.}~\bibnamefont
  {Merlo}}, \bibinfo {author} {\bibfnamefont {J.}~\bibnamefont {Dominski}},
  \bibinfo {author} {\bibfnamefont {A.}~\bibnamefont {Bhattacharjee}}, \bibinfo
  {author} {\bibfnamefont {C.}~\bibnamefont {Chang}}, \bibinfo {author}
  {\bibfnamefont {F.}~\bibnamefont {Jenko}}, \bibinfo {author} {\bibfnamefont
  {S.}~\bibnamefont {Ku}}, \bibinfo {author} {\bibfnamefont {E.}~\bibnamefont
  {Lanti}}, \ and\ \bibinfo {author} {\bibfnamefont {S.}~\bibnamefont
  {Parker}},\ }\href@noop {} {\bibfield  {journal} {\bibinfo  {journal} {Phys.
  Plasma}\ }\textbf {\bibinfo {volume} {25}},\ \bibinfo {pages} {062308}
  (\bibinfo {year} {2018})}\BibitemShut {NoStop}%
\bibitem [{\citenamefont {Burckel}\ \emph {et~al.}(2010)\citenamefont
  {Burckel}, \citenamefont {Sauter}, \citenamefont {Angioni}, \citenamefont
  {Candy}, \citenamefont {Fable},\ and\ \citenamefont
  {Lapillonne}}]{burckel10}%
  \BibitemOpen
  \bibfield  {author} {\bibinfo {author} {\bibfnamefont {A.}~\bibnamefont
  {Burckel}}, \bibinfo {author} {\bibfnamefont {O.}~\bibnamefont {Sauter}},
  \bibinfo {author} {\bibfnamefont {C.}~\bibnamefont {Angioni}}, \bibinfo
  {author} {\bibfnamefont {J.}~\bibnamefont {Candy}}, \bibinfo {author}
  {\bibfnamefont {E.}~\bibnamefont {Fable}}, \ and\ \bibinfo {author}
  {\bibfnamefont {X.}~\bibnamefont {Lapillonne}},\ }\href@noop {} {\bibfield
  {journal} {\bibinfo  {journal} {J. Phys.: Conf. Ser.}\ }\textbf {\bibinfo
  {volume} {260}},\ \bibinfo {pages} {012006} (\bibinfo {year}
  {2010})}\BibitemShut {NoStop}%
\bibitem [{\citenamefont {Moritaka}\ and\ \citenamefont
  {et~al.}(tion)}]{moritaka_prep}%
  \BibitemOpen
  \bibfield  {author} {\bibinfo {author} {\bibfnamefont {T.}~\bibnamefont
  {Moritaka}}\ and\ \bibinfo {author} {\bibnamefont {et~al.}},\ }\href@noop {}
  {\  (\bibinfo {year} {In preparation})}\BibitemShut {NoStop}%
\end{thebibliography}%

\end{document}